\documentclass[conference]{IEEEtran}
\IEEEoverridecommandlockouts
\usepackage{cite}
\usepackage{amsmath,amssymb,amsfonts}
\usepackage{algorithmic}
\usepackage{graphicx}
\usepackage{textcomp}
\usepackage{xcolor}
\def\BibTeX{{\rm B\kern-.05em{\sc i\kern-.025em b}\kern-.08em
		T\kern-.1667em\lower.7ex\hbox{E}\kern-.125emX}}

\usepackage[left=0.65in,right=0.65in,bottom=0.993in,top=0.72in]{geometry}

\usepackage{amsmath,amssymb,amsfonts,amsthm}
\usepackage{amsmath}
\usepackage[caption=false,font=footnotesize,labelfont=rm,textfont=rm]{subfig}
\usepackage{graphicx}
\usepackage{bbm}

\newcommand{\red}[1]{\color{red}#1}
\usepackage[colorlinks=true,linkcolor=red,citecolor=green]{hyperref}

\usepackage{algorithmic,algorithm}
\usepackage{setspace}

\usepackage{float}
\usepackage{subfig}
\usepackage{overpic}
\usepackage{epstopdf}
\usepackage{cases}
\usepackage{bm}

\usepackage{stfloats} 

\allowdisplaybreaks[4]

\usepackage{makecell} 

\begin{document}
	\title{Semantic Noise Aided Secure Image \\ Transmission over MIMO Fading Channels}
	\author{	
		\IEEEauthorblockN{Xue Han, Biqian Feng, Ting Zhou, Yongpeng Wu, \emph{Senior Member, IEEE}, Yuanwei Liu, \emph{Fellow, IEEE}, \\ Arumugam Nallanathan, \emph{Fellow, IEEE}, Xiang-Gen Xia, \emph{Fellow, IEEE}, and Wenjun Zhang, \emph{Fellow, IEEE}}


		\thanks{Xue Han, Biqian Feng, Yongpeng Wu, and Wenjun Zhang are with the School of Integrated Circuits (School of Information Science and Electronic Engineering), Shanghai Jiao Tong University, Shanghai 200240, China (e-mail: han.xue@sjtu.edu.cn; fengbiqian@sjtu.edu.cn; yongpeng.wu@sjtu.edu.cn; zhangwenjun@sjtu.edu.cn).}
		
		\thanks{Ting Zhou is with the School of Microelectronics, Shanghai University, Shanghai 201210, China, and also with Shanghai Frontier Innovation Research Institute, Shanghai 201100, China (e-mail: zhouting@shu.edu.cn).}
		
		\thanks{Yuanwei Liu is with the Department of Electrical and Electronic Engineer-
			ing, the University of Hong Kong, and also with Faculty of Applied Sciences,
			Macao Polytechnic University (e-mail: yuanwei@hku.hk).}
		
		\thanks{Arumugam Nallanathan is with the School of Electronic Engineering and Computer Science, Queen Mary University of London, London and also with the Department of Electronic Engineering, Kyung Hee University, Yongin-si, Gyeonggi-do 17104, Korea (mailto:a.nallanathan@qmul.ac.uk).}

		\thanks{Xiang-Gen Xia is with the Department of Electrical and Computer Engineering, University of Delaware, Newark, DE 19716, USA (e-mail: xxia@ee.udel.edu).}
	}
	\maketitle

	\begin{abstract}
		Existing semantic communications have exhibited satisfactory performance in many tasks, but secure image transmission remains insufficiently explored. We propose a novel secure image semantic communication (SISC) framework over multiple-input multiple-output (MIMO) fading channels. To ensure high-quality image reconstruction for the legitimate semantic user (SU) and simultaneously interfere with the eavesdropper (Eve), we design a semantic noise generation (SNG) network. This network generates a beneficial semantic noise map based on both the source features and the SU channel state information (CSI). An efficient channel estimation enhanced network is incorporated to obtain the accurate CSI and enhance the system performance. Furthermore, to improve the secure image reconstruction quality, we develop an efficient transceiver beamformer optimization algorithm, where the formulated problem is solved using the constrained stochastic successive convex approximation method. In the proposed SISC framework, semantic noise generation and beamforming optimization work together to ensure secure and high-quality image transmission.
		Numerical results demonstrate that the proposed semantic noise aided transmission scheme effectively protects image information from leakage to Eve while maintaining high-fidelity image reconstruction at SU.
	\end{abstract}


	\section{Introduction}
	As a promising technology for sixth-generation (6G) wireless communications, semantic communication aims to enhance transmission efficiency by conveying only task-relevant information, thereby enabling a deeper integration between information and communication technologies (ICT) and artificial intelligence (AI) \cite{ping_sem}\cite{Han_SCSC}. Despite its great potential, the inherent openness of wireless semantic communication exposes semantic-rich data to significant security and privacy threats, rendering it vulnerable to eavesdropping attacks. Furthermore, since semantic information typically encapsulates more sensitive content than raw data, stricter security mechanisms are essential to ensure data confidentiality and integrity over wireless channels \cite{weixuan chen}, especially when implemented on resource-constrained devices.

	Physical layer security (PLS) is an efficient technology for achieving secure communications in wireless networks by exploiting the inherent randomness and dynamics of wireless channels to ensure data confidentiality \cite{security}.
	Mukherjee et al. \cite{Mukherjee} developed two low-complexity algorithms with closed-form iterative updates to efficiently compute the secrecy capacity of multiple-input multiple-output (MIMO) wiretap channels.
	In \cite{Zhang2023}, the authors proposed optimized hybrid beamforming algorithms to enhance the secrecy capacity of near-field MIMO downlink channels and demonstrated the inherent security benefits of operating in the near-field region.
	An et al. \cite{An_RIS} proposed a robust beamforming method to combat large-scale fading and enable simultaneous information and power transfer.
	Moreover, Lin et al. investigated a joint beamforming and phase-shift optimization scheme to improve link reliability \cite{Zhi_2} and designed dedicated beamforming strategies to enhance physical-layer secrecy against eavesdropping \cite{Zhi_3}.
	The artificial noise (AN) technology has been widely investigated as a typical PLS
	technique due to its ability to utilize channel state information (CSI) to enhance the secrecy performance. By generating the AN in the null space of the legitimate channel, the interference affects only the eavesdroppers, thereby enhancing the secrecy performance \cite{Choi2022}.
	For example, Zhao et al. \cite{Zhao2018} proposed an AN-aided interference alignment framework that considers energy harvesting, thereby meeting both security and energy efficiency requirements in information transmission.
	Yin et al. \cite{Yin_2022} investigated a green interference-based symbiotic security scheme that exploits co-channel and inter-beam interference to maximize the secrecy rate in integrated satellite terrestrial networks.
	In \cite{Yin_2023}, the authors studied secure transmission in heterogeneous space–air–ground integrated networks and proposed a digital-twin-assisted multi-point symbiotic security scheme.
	Yun et al. \cite{Yun2020} proposed a deep neural network (DNN)-based approach, which jointly optimizes the precoders of information signal and AN. While traditional solutions like encryption can protect privacy, they often incur high computational and communication costs, leading to reduced transmission efficiency. Moreover, traditional physical layer security schemes primarily focus on symbol- or signal-level protection and thus fail to address the inherent security and privacy issues arising in semantic communication.
	
	In recent years, increasing attention has been paid to security concerns in semantic communication scenarios. 
	Du et al. \cite{3} discussed the security challenges of wireless communications for the semantic Internet of Things and examined the applicability of conventional security techniques to semantic communications, including physical-layer security, covert transmission, and encryption. The potential privacy and security vulnerabilities in semantic communications were further investigated in \cite{4}.
	In the presence of malicious jamming, work \cite{5} proposed defense mechanisms based on weighted perturbations to enhance robustness against adversarial attacks.
	In recent studies, information bottleneck and adversarial learning techniques were incorporated into semantic communication frameworks to strengthen privacy protection \cite{Wang_IB}.
	Deep learning (DL)-based semantic communications with joint source and channel coding (JSCC) have emerged as a promising solution. Tung et al. \cite{DeepJSCEC} proposed the first Deep JSCC-based framework for wireless image transmission that achieves robustness against eavesdropping without relying on assumptions about the eavesdropper’s channel or decoding intent.
	Erdemir et al. \cite{Erdemir} formulated an adversarial training framework to jointly optimize semantic encoding and decoding, achieving a balance between minimizing information leakage and maintaining reconstruction quality.
	Zhang et al. \cite{Maojun Zhang} developed a joint source–channel autoencoder for image semantic extraction, employing a mean squared error (MSE) loss function to effectively balance transmission efficiency and privacy protection.
	While improving the security and robustness of systems, the communication overhead and the impact of transmission efficiency should be carefully considered.
	Li et al. \cite{Yongkang Li} introduced a deep neural network (DNN)-based secure semantic communication framework, named DeepSSC, which adopts a two-stage training strategy to jointly balance transmission reliability and security by integrating physical layer security and variational inference principles. 
	Due to the broadcast nature of wireless channels \cite{Chen2024}, eavesdroppers can intercept transmitted signals and exploit a leaked semantic decoder to infer the original information content.
	To address this, Luo et al. \cite{Xinlai Luo} introduced a secret key-based encryption mechanism combined with adversarial encryption training to preserve security while ensuring accurate semantic recovery. 
	To satisfy diverse semantic protection requirements and achieve the objectives of secrecy, privacy, and integrity, an online semantic communication framework with three hot-pluggable modules was proposed in \cite{Semiprotector}.

	Although numerous methods have been developed for secure data transmission, the potential of exploiting semantic noise in wireless communications has not been thoroughly investigated, particularly with respect to leveraging the characteristics of the wireless channel environment.
	In contrast to technical noise caused by physical impairments such as channel noise, interference, or fading, semantic noise originates from problems in understanding the context, mismatches in knowledge, or flaws in the deep learning models employed for semantic encoding and decoding \cite{SN2025}.
	Recently, Hu et al. \cite{Qiyu Hu} proposed a robust semantic communication framework by generating semantic-noise-augmented training samples to enhance system resilience. Mu et al. \cite{Xidong Mu} demonstrated that the semantic signal can inherently act as beneficial artificial noise, which not only remains confidential to the eavesdropper but also interferes with it, thereby improving secrecy performance. Existing physical layer security schemes primarily focus on symbol- or signal-level protection and thus fail to address the inherent security and privacy issues arising in semantic communication.

	In this work, we go one step further to explore the potential of semantic communication in PLS. 
	In particular, we propose a novel semantic noise-aided secure image communication framework (SISC) over MIMO fading wiretap channels, where the transmitter aims to send image source data to the legitimate user in the presence of an eavesdropper. 
Different from existing semantic security approaches that apply pre-defined perturbations and AN methods that inject power-domain interference, the proposed SISC generates a beneficial semantic noise map based on both the source image features and the CSI of the legitimate user, enabling dynamic and channel-aware security enhancement.
	We adopt the Swin Transformer \cite{SwinTF} as the backbone and further investigate the internal attention mechanism within the semantic encoder. A semantic noise-aware multi-head self-attention (SN-MSA) module is proposed to effectively protect semantic features by jointly considering the CSI and the semantic noise map. Specifically, to acquire accurate legitimate CSI, we propose a channel enhancement estimator in the SISC framework for combating channel variation. In order to obtain the suitable semantic noise map, a learnable semantic noise generator is introduced as a preprocessing stage, which enables the elements in attention weights to adaptively interfere according to source semantic features and legitimate channel conditions. 
	Moreover, we design a transceiver beamforming optimization algorithm that effectively enhances secrecy performance, thereby maximizing confidential information delivery while suppressing information leakage to the eavesdropper.
	The main contributions of this paper can be summarized as follows.
	
	\begin{itemize}
		\item {\textit{SISC Framework:} We propose SISC, a semantic noise-aided secure image semantic communication framework for MIMO fading wiretap channels, which achieves high reconstruction accuracy at the legitimate semantic user (SU) while degrading the semantic inference capability of the eavesdropper (Eve). In this framework, the proposed dual CSI-semantic noise (CSI-SN) aware SN-MSA network incorporates the legitimate CSI together with appropriately generated semantic noise to improve transmission security and mitigate information leakage.}
		
		\item {\textit{Channel Estimation and Semantic Noise Generation:}} To ensure accurate CSI acquisition and strengthen the robustness of semantic transmission, we develop a channel estimation enhanced network (CEEN) that estimates legitimate CSI under fading conditions. Furthermore, a semantic noise generation (SNG) network is introduced to produce task-relevant semantic noise maps, which are incorporated into the semantic encoder to improve security and effectively confuse the eavesdropper.

		\item {\textit{Beamforming Optimization Algorithm Design:}} We further design a constrained successive convex approximation (CSSCA)-based algorithm to solve the transceiver beamformer design problem in the MIMO wiretap scenario. It guarantees secure semantic transmission even under the worst-case eavesdropping condition, where Eve adopts the optimal linear combiner.
		
		\item {\textit{Performance Validation:} Simulation results demonstrate that the legitimate SU in the proposed SISC scheme achieves superior reconstruction performance compared with existing baselines, while the eavesdropper Eve experiences severely degraded recovery quality, highlighting the effectiveness and security advantages of the proposed approach.}	
		
	\end{itemize}
	
	\begin{figure}
	\centering
	\subfloat[]{
		\label{System}
		\includegraphics[width=0.93\linewidth]{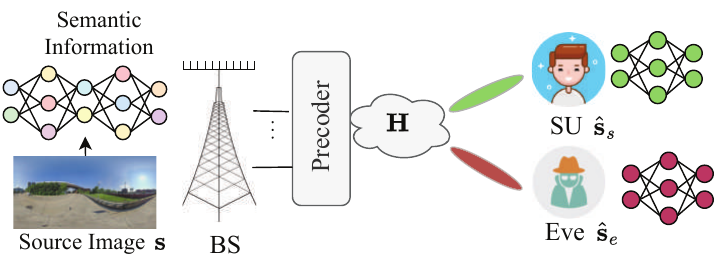}
	}\hfill
	\subfloat[]{
		\label{Network}
		\includegraphics[width=0.93\linewidth]{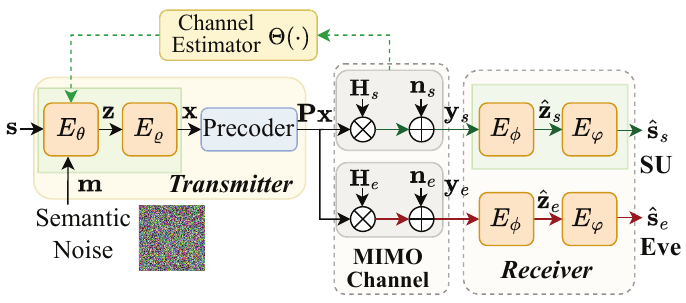}
	}
	\caption{(a) System model of semantic image communication network. (b) The framework of the semantic communication system with semantic noise for image transmission over MIMO wiretap channels.}
	\label{system}
	\vspace{-0.2cm}
\end{figure}

	The remainder of the paper is structured as follows. Section \ref{sec:system model} introduces the considered system model over wiretap channels. Section \ref{sec:SISC} describes the detailed SISC framework structures. Section \ref{sec:SN} provides the generation process of learnable semantic noise.
	Section \ref{sec:Transceiver} introduces an efficient transceiver beamforming optimization algorithm. 
	Section \ref{sec:Results} presents the numerical results and performance evaluation. 
	Finally, Section \ref{sec:Conclusion} concludes this paper.
	
	\textit{Notations}: 
	Superscripts $T$, $H$, and $\dag$ stand for the transpose, conjugate transpose, and pseudo inverse, respectively. $ \mathbb{R} $ and $\mathbb{C} $ denote the sets of real and complex numbers, respectively. $\| \cdot \|$ represents the Frobenius norm. $\mathcal{CN}(\mu,\sigma^2)$ denotes the circularly symmetric complex Gaussian (CSCG) distribution with mean $\mu$ and variance $\sigma^2$.

	\section{System Model}
	\label{sec:system model}

	In this section, we first present the system model of secure image semantic communication (SISC) over MIMO wiretap channels. Then, an objective loss function is formulated to characterize the trade-off between accurate image recovery at the semantic user (SU) and the information leakage to the eavesdropper (Eve).

	\subsection{Secure Semantic Transmission System}
	We consider wireless semantic image transmission over MIMO block fading channels as demonstrated in Fig. \ref{system}\subref{System}. At the transmitter (Tx), the base station (BS), which is equipped with $N_t$ transmit antennas, aims to securely transmit images to the legitimate SU equipped with $N_r$ antennas while operating in the presence of Eve with $N_e$ antennas at the receiver (Rx). 
	
	\subsubsection{Semantic-aware Transmitter}
	Let $\mathbf s \in \mathbb R^{H \times W \times 3}$ denote the source image transmitted over a wiretap channel to a legitimate SU for recovery, where $H$, $W$, and 3 denote the height, width, and the number of color channels in the RGB format of a source image, respectively. 
	As shown in Fig. \ref{system}\subref{Network}, 
	at the BS side, the semantic encoder $E_{\theta}(\cdot): \mathbb R^{H \times W \times 3} \mapsto \mathbb R^{C_L}$ parameterized by $\theta$, encodes the source image with the help of side information including estimated channel state information (CSI) of SU, $\hat{\mathbf{H}}_s \in \mathbb C^{N_r \times N_t}$, and the semantic noise, $\mathbf m$, into the semantic feature $\mathbf z \in \mathbb R^{C_L}$, where $C_L$ refers to the sequence length of the extracted semantic information. Then, the channel encoder, $E_{\varrho}(\cdot): \mathbb R^{C_L} \mapsto \mathbb C^{k}$ parameterized by $\varrho$, maps the feature $\mathbf z$ into a complex-valued vector $\mathbf{x} \in \mathbb C^{k}$ to combat channel fading and noise, i.e., $\mathbf{x} = E_{\varrho}(\mathbf z)$, where $k$ denotes the length of the transmitted semantic codeword sequence. The channel bandwidth ratio (CBR) is defined as $\rho = \frac{k}{H \times W \times 3}$, which reflects the utilization of channel resources per symbol.
	During this process, we can also get the mean value $\nu$ and variance value $\kappa$ of the input sample data $\mathbf{s}$. 
	The transmitted signal $\mathbf x$ satisfies the average power constraint
	\begin{equation}
		\frac{1}{N_tk}\| \mathbf x\|^2 \leq 1.
		\label{power_x}
	\end{equation}
	The whole semantic information encoding process can be summarized as
	\begin{equation}
		\mathbf{s} \xrightarrow{E_{\theta}(\cdot)}
		\mathbf z \xrightarrow{E_{\varrho}(\cdot)}
		{\mathbf{x}},
	\end{equation}
	where the semantic symbol vector ${\mathbf{x}}$ is given by
	\begin{equation}
		\mathbf{x} = {E_{\varrho}(E_{\theta}(\mathbf{s}, \mathbf{H}_s, \mathbf m))}.
	\end{equation}

	\subsubsection{Wireless Wiretap Channel Transmission}
	After DeepJSCC encoder mapping, the BS adopts the transmit beamformer $\mathbf P \in \mathbb C^{N_t \times k}$ to transmit the signal data. Considering a linear channel model, the signals received by SU $\mathbf{y}_s$ and the eavesdropper Eve $\mathbf{y}_e$ are given by
	\begin{equation}
		\mathbf{y}_s = \mathbf{H}^H_s \mathbf{P}\mathbf{x} +\mathbf{n}_s, 
		\label{y_s}
	\end{equation}
	\begin{equation}
		\mathbf{y}_e = \mathbf{H}^H_e \mathbf{P}\mathbf{x} +\mathbf{n}_e,
	\end{equation}
	respectively, where $\mathbf{H}_s \in \mathbb C ^{N_r \times N_t}$ and $\mathbf{H}_e \in \mathbb C ^{N_e \times N_t}$ denote the channel matrices from the perspectives of SU and Eve, respectively. $\mathbf{n}_s \sim \mathcal{CN}(0,\sigma_s^2 \mathbf I)$ and $\mathbf{n}_e \sim \mathcal{CN}(0,\sigma_e^2 \mathbf I)$ denote complex independent and identically distributed (i.i.d.) channel noise vectors. Specifically, to acquire the accurate CSI, we transmit pieces of known pilots, $\mathbf A \in \mathbb C^{N_t \times N_t}$. Both the transmitted and received pilot signals are exploited to estimate the CSI. The received pilot can be expressed as
	\begin{equation}
		\hat{\mathbf A} = {\mathbf{H}}_s \mathbf A+ \mathbf{n},
	\end{equation}
	where $\hat{\mathbf A} \in \mathbb R^{N_r \times N_t}$ denotes the received pilots at SU, $\mathbf{n}$ is the additive channel noise experienced during the pilot transmission. 
	The estimated CSI matrix $\hat{\mathbf{H}}_s$ can be given by
	\begin{equation}
		\hat{\mathbf{H}}_s = \Theta({\mathbf A},	\hat{\mathbf A}),
	\end{equation}
	where $\Theta(\cdot)$ is the channel estimator. 
	Let $P_\mathrm{max}$ denote the power budget, and the transmit signal satisfies the power constraint, 
	\begin{equation}
		\mathrm{Tr}(\mathbf{P}\mathbf{P}^H) \leq P_\mathrm{max}.
		\label{power_con}
	\end{equation}
	
	\begin{figure*}
		\centering
		\vspace{-0.04cm} 
		\setlength{\abovecaptionskip}{0cm} 
		\setlength{\belowcaptionskip}{-1.84em} 
		\includegraphics[width=0.95\linewidth]{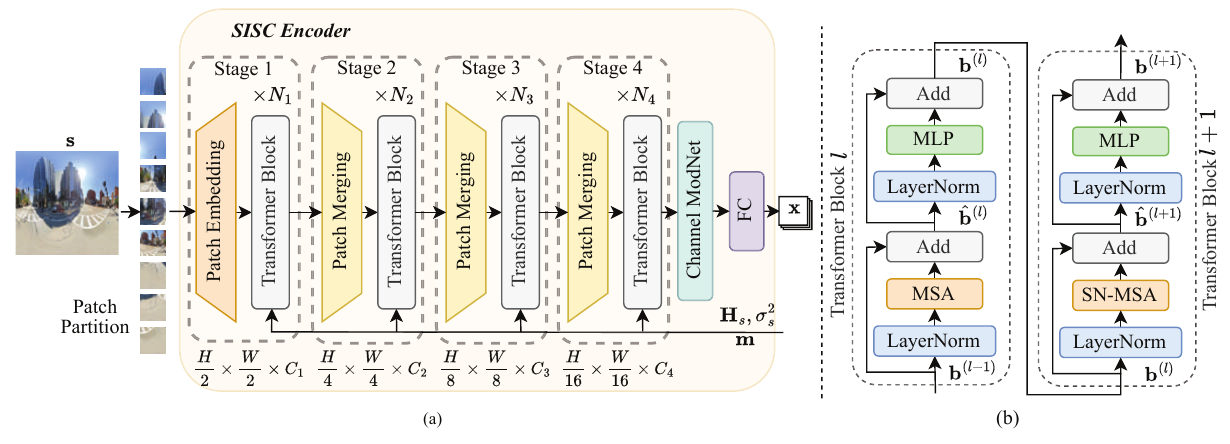}
		\caption{(a) The detailed structure of the SISC semantic encoder network. (b) Two consecutive Swin Transformer blocks. }
		\label{DJSCC}
		\vspace{-0.25cm}
	\end{figure*}

	\subsubsection{Semantic-aware Receiver}
	Consider the linear receive beamforming strategy so that the estimated signal at the legitimate SU is given by
	\begin{equation}
		\bar {\mathbf y}_s = \mathbf {U}_s^H {\mathbf y}_s, 
	\end{equation}
	where $\mathbf {U}_s \in \mathbb C^{N_r \times N_r}$
	is a receive beamformer for legitimate SU to detect ${\mathbf x}$.
	Similarly, Eve also employs a receive beamformer $\mathbf {U}_e^H \in \mathbb C^{N_e \times N_e}$	to enhance its interception capability
	\begin{equation}
		\bar{\mathbf{y}}_e = \mathbf{U}_e^H {\mathbf y}_e.
	\end{equation}
	At the eavesdropper receiver, the semantic decoder employs the same network architecture and shares weight parameters as those of the legitimate SU to ensure a fair and worst-case performance comparison. We assume Eve employs the optimal linear combiner to maximize its recovered signal quality. Taking the MMSE receiver as an example, the optimal linear combiner $\mathbf{U}_e^\mathrm{opt}$ at Eve side can be described as \cite{MMSE}
	\begin{equation}
		\mathbf{U}_e^{\mathrm{opt}}=\left(\mathbf{H}_e\mathbf{P}\mathbf{P}^H\mathbf{H}_e^H+\sigma_e^2\mathbf{I}_{N_e}\right)^{-1}\mathbf{H}_e\mathbf{P}.
	\end{equation}
	The channel matrix of Eve, i.e., $\mathbf{H}_e \in \mathbb C ^{N_e \times N_t}$, is modeled as an i.i.d. Rayleigh fading matrix. 
	Furthermore, the transmission rate $R_s$ (bps/Hz) at user SU is given by
	\begin{equation}
		R_{s}\triangleq \log\det\left(\mathbf{I}+ \frac{1}{\sigma_{s}^{2}} \mathbf{U}_{s}^H\mathbf{H}_{s}\mathbf{P}\mathbf{P}^H\mathbf{H}_{s}^H \mathbf{U}_{s}\right).
	\end{equation}
	Benefiting from the semantic communication framework, the transmission delay $T_s$ of the $i$-th source image at the legitimate SU is mapped by
	\begin{equation}
		T_s = \frac{k}{BR_s},
		\label{delay}
	\end{equation}
	where $B$ (Hz) is the bandwidth in the system.
	Given the maximum transmission delay $T_\mathrm{max}$, the maximum transmission delay constraint is equivalent to
	\begin{equation}
		R_s \ge R_s^\mathrm{min} \triangleq \frac{k}{B T_\mathrm{max}}.
	\end{equation}

	For the legitimate Rx, the received semantics pass through the channel decoder, $E_{\phi}(\cdot): \mathbb{C}^{k} \mapsto \mathbb R^{C_L}$ parameterized by $\phi$ to obtain semantics $\hat {\mathbf z}_s \in \mathbb R^{C_L}$. Then the semantic decoder, $E_{\varphi}(\cdot): \mathbb R^{C_L} \mapsto \mathbb R^{H \times W \times 3}$ with parameter $\varphi$, reconstructs the source image $\hat{\mathbf{s}}_s$. Meanwhile, eavesdropper Eve attempts to reconstruct the transmitted image from the received codewords into the image $\hat{\mathbf{s}}_e$. The image reconstruction process at the legitimate SU can be expressed by
	\begin{equation}
		\bar{\mathbf{y}}_s \xrightarrow{E_{\phi}(\cdot)}
		\hat{\mathbf z}_s \xrightarrow{E_{\varphi}(\cdot)}
		{\hat{\mathbf{s}}_s},
	\end{equation}
	where the recovered $\hat{\mathbf{s}}_s$ is given by
	\begin{equation}
		\hat{\mathbf{s}}_s = {E_{\varphi}(E_{\phi}(\bar{\mathbf{y}}_s))}.
	\end{equation}
	
	At the eavesdropper Rx, the semantic decoder and channel decoder at Eve employ the same network architecture and shared weight parameters as those of the legitimate SU to ensure a fair performance comparison. 
	Thus, this image eavesdropping process can be represented as
	\begin{equation}
		\bar{\mathbf{y}}_e \xrightarrow{E_{\phi}(\cdot)}
		\hat {\mathbf z}_e \xrightarrow{E_{\varphi}(\cdot)}
		{\hat{\mathbf{s}}_e},
	\end{equation}
	where $\hat{\mathbf{s}}_e \in \mathbb R^{H \times W \times 3}$ and $\hat {\mathbf z}_e \in \mathbb R^{C_L}$ represent the reconstructed image and recovered semantics at Eve, respectively.
	
	\subsection{Objective Function Formulation}
	To measure the reconstruction loss, we adopt the mean squared error (MSE) loss between the recovered and the original image for network training, which is defined as 
	\begin{equation}
		\begin{aligned}
			&\mathcal{L}_s = \frac{1}{N}\sum\limits_{i=1}^{N}\|\hat{\mathbf s}_{s}(i) - \mathbf s(i)\|^2, \\
			&\mathcal{L}_e = \frac{1}{N}\sum\limits_{i=1}^{N}\|\hat{\mathbf s}_{e}(i) - \mathbf s(i)\|^2,
		\end{aligned}
	\end{equation}
	where $N$ is the total number of training samples, $\mathbf s(i)$ is the $i$-th transmitted image, $\hat{\mathbf s}_{s}(i)$ and $\hat{\mathbf s}_{e}(i)$ are the $i$-th recovered images of SU and Eve, respectively. We aim to simultaneously minimize the MSE at the legitimate SU and information leakage to the eavesdropper Eve by concurrently optimizing the transmit beamformer and the semantic noise.
	The corresponding loss function is formulated as
	\begin{equation}
		\mathcal{L}(\mathbf{P}, \mathbf{U}_s, \mathbf m) \triangleq \mathcal{L}_s - \lambda \mathcal{L}_e,
		\label{p: initial}
	\end{equation}
	where $\lambda\ge0$ denotes the trade-off parameter. A larger $\lambda$ emphasizes information protection by penalizing Eve’s reconstruction capability.

	\section{Detailed Structure of SISC}
	\label{sec:SISC}
	In this section, we present the details and key methodologies of the proposed secure image semantic communication (SISC) framework, including the overall semantic encoder network structure, the dual CSI and semantic noise (CSI-SN) aware SN-MSA network, the channel estimation module CEEN, and other SISC network structures.
	
	\subsection{The Overall Architecture of SISC}
	An overview of the proposed semantic encoder, which is built upon the Swin Transformer backbone \cite{SwinTF}, is shown in Fig. \ref{DJSCC}{\red{(a)}}. The source image is partitioned into $\frac{H}{2} \times \frac{W}{2}$ non-overlapping patches to form the input semantic image sequence arranged in a left-to-right and top-to-bottom order. Then, the sequences are fed into a series of fully connected layers for linear projection and feature flattening to generate feature embedding tokens. Subsequently, the feature embeddings are processed through a four-stage hierarchical structure. Each stage is encapsulated by a patch merging module and $N_j$ transformer blocks $ (j = 1, 2, 3, 4)$. These blocks aim to downsample the feature embedding tokens into $\frac{H}{2^j} \times \frac{W}{2^j} \times C_j$, where $C_j$ denotes the embedding dimension. The estimated legitimate CSI and the semantic noise are incorporated as side information into the transformer block to enhance semantic feature extraction.
	Fig. \ref{DJSCC}{\red{(b)}} illustrates the structure of two successive transformer blocks, each consisting of a layer normalization, an attention layer, and a multi-layer perceptron (MLP), yet differing in the architecture of attention modules.

	\begin{figure}
		\centering
		\setlength{\belowcaptionskip}{-4.84em} 
		\includegraphics[width=0.92\linewidth]{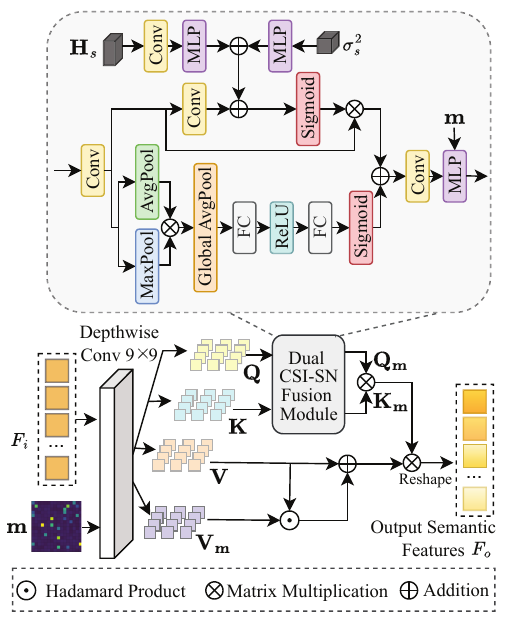}
		\caption{The SN-MSA network and the detailed architecture of the dual CSI-SN fusion module.}
		\label{MHA_CSI+SN}
		\vspace{-0.15cm}
	\end{figure}

	\subsection{Dual CSI-SN Aware SN-MSA Network}
	\subsubsection{The SN-MSA Structure} 
	The semantic noise-aided multi-head self-attention (SN-MSA) network is developed based on the standard multi-head self-attention (MSA) structure. The self-attention mechanism allows the model to capture long-range dependencies within the image.
	As shown in Fig. \ref{MHA_CSI+SN}, the input semantic sequence $F_i$ is projected into three attention matrices by a depthwise separable convolution, namely the query matrix $\mathbf{Q} \in \mathbb{R}^{C_L \times C_D}$, the key matrix $\mathbf{K} \in \mathbb{R}^{C_L \times C_D}$, and the value matrix $\mathbf{V} \in \mathbb{R}^{C_L \times C_D}$, where $C_L$ is the sequence length of the extracted semantic information and $C_D$ denotes the number of patch embeddings. 
	The value matrix $\mathbf{V}$ characterizes the importance of each element in the input sequence, while the query $\mathbf{Q}$ and $\mathbf{K}$ matrices together model the correlations between two arbitrary input elements in the sequence. At the same time, the value matrix $\mathbf{V_m} \in \mathbb{R}^{C_L \times C_D}$ of the semantic noise map is obtained by applying the same depthwise separable convolution to the semantic noise map by weight-sharing, producing semantic noise-aware value features. To facilitate accurate codeword transmission and effective semantic extraction for the SU while disrupting Eve, we design a dual CSI-SN fusion network to leverage the beneficial semantic noise map and the feedback CSI to refine the distribution of attention weights. 
	
	\subsubsection{The Dual CSI-SN Fusion Module}
	The CSI-SN fusion module leverages a dual-branch architecture to jointly consider semantic feature granularity, legitimate channel state adaptation, and the semantic noise map. In Fig. \ref{MHA_CSI+SN}, the input features first pass through a convolution layer, producing two identical feature maps, which are then fed into two parallel branches. 
	The two branches play distinct functional roles.
	i) In the upper branch, the estimated CSI $\mathbf H_s$ and $\sigma_s^2$ are first processed to obtain the corresponding channel features \cite{H}. 
	This branch focuses on physical layer adaptation, generating channel-wise attention weights to reweight the semantic features $F_i$ based on the current channel quality.
	ii) The lower branch applies both max-pooling and average-pooling operations to capture local details and global context. 
	This branch focuses on content importance to obtain semantic importance weights.
	The pooled features are then processed by a global average pooling layer to obtain compact representations. 
	These representations are passed through two fully connected layers and a sigmoid activation function to derive the semantic importance weights of each input patch, which guide the subsequent CSI fusion. 
	Finally, the outputs of the two branches, containing the refined semantic importance and channel features, respectively, are concatenated with the semantic noise map $\mathbf m \in \mathbb R^{C_L \times C_L}$, where $C_L\times C_L$ is the $\mathbf{Q} \mathbf{K}^T$ attention weights dimension.
	In this proposed dual CSI-SN fusion module, the attention weight is generated based on the computed feature semantic importance along with the estimated CSI from the legitimate channel. 
	Thus, after the feature embeddings $\mathbf{Q}$ and $\mathbf{K}$ are combined with the semantic noise map, we can obtain 
	\begin{equation}
		\mathbf{Q_m} \mathbf{K}^T_\mathbf{m}=\mathbf m \odot (\mathbf{Q} \mathbf{K}^T),
	\end{equation}
	where $\odot$ is the Hadamard product.

\begin{figure*}[t]
	\centering
	\setlength{\belowcaptionskip}{-2.84em} 
	\includegraphics[width=0.95\linewidth]{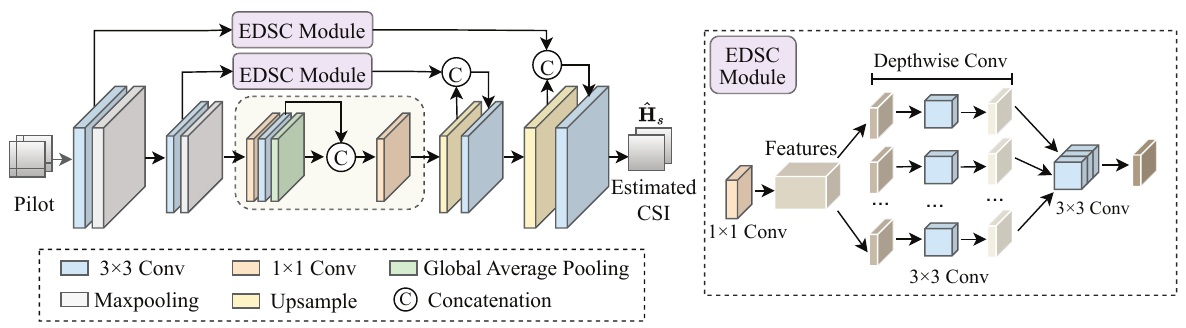}
	\caption{The architecture of the channel estimation enhanced network (CEEN) and the enhanced depthwise separable convolution (EDSC) network.}
	\label{Unet_CE}
\end{figure*}

	\subsubsection{Attention Score}
	Given the flattened feature map $\mathbf b^{(l)}$, the query, key and value matrices $\mathbf{Q}$, $\mathbf{K}$ and $\mathbf{V}$ are computed as
	\begin{equation}
		\mathbf{Q}= \mathbf b^{(l)} \mathbf {W_Q}, \mathbf{K}= \mathbf b^{(l)} \mathbf {W_K}, \mathbf{V}= \mathbf b^{(l)} \mathbf {W_V},
	\end{equation}
	where $\mathbf {W_Q} , \mathbf {W_K}, \mathbf {W_V} \in \mathbb R^{C_D \times C_D}$ are learnable projection matrices. $l$ denotes the $l$-th transformer block.
	In this way, the self-attention score of SN-MSA can be computed as
	\begin{equation}
		\label{attention_m}
		\begin{aligned}
			\textrm{Attention} (\mathbf{Q_m}, \mathbf{K_m}, \mathbf{V}, \mathbf{V_m}) = &\textrm{Softmax}(\frac{\mathbf{Q_m} \mathbf{K}_{\mathbf{m}}^T}{\sqrt{d}}) \\
			&\otimes (\mathbf{V} \oplus \mathbf{V} \odot \mathbf{V_m}),
		\end{aligned}
	\end{equation}
	where $d$ represents the constant scaling parameter to adjust the attention weight of $\mathbf{Q_m} \mathbf{K}_{\mathbf{m}}^T$.
	The embedding $\mathbf{V}$ remains unchanged, as it is derived only from the original semantic feature sequences. Finally, we can obtain the output semantic features $F_o$.
	
	Based on the modified attention scores in Eq. \eqref{attention_m}, the dual CSI-SN fusion module in the SN-MSA network improves attention distributions under varying channel conditions for SU. It selectively perturbs less salient elements in $\mathbf{Q}$ and $\mathbf{K}$ under strong interference and eavesdropping, while preserving semantically important information in $\mathbf{V}$, thus maintaining robust feature representation for legitimate SU.

\begin{algorithm}[t]
	\caption{Pretraining Strategy of CEEN.}
	\label{Alg_H}
	{\bf Input:} Sample set of MIMO Channel fading matrices $\mathcal{H}$, and pilot sequence $\mathbf A$.\\
	{\bf Output:} The trained channel estimator $\Theta(\cdot)$.
	
	\begin{algorithmic}[1]
		\STATE Sampling: Select a batch $\mathbf{B} \in \mathcal{H}$.
		\STATE Transmission: Transmit the stable pilot sequence $\mathbf A$ through the generated MIMO fading channels.
		\STATE Estimation: Feed $\mathbf A$ and received $\hat {\mathbf A}$ into the channel estimator $\Theta(\cdot)$ to obtain the estimated CSI $\hat{\mathbf{H}}_{s}$, i.e., $\hat{\mathbf{H}}_{s}=\Theta(\mathbf A, \hat {\mathbf A})$.
		\STATE Optimization: Compute the loss using \eqref{H_opt}.
		\STATE Update the network parameters via gradient descent and return the trained estimator $\Theta(\cdot)$.
	\end{algorithmic}
\end{algorithm}

	\subsection{Channel Estimation Enhanced Network (CEEN)}
	To acquire accurate feedback CSI of the legitimate SU in the SISC framework, we propose a channel estimation enhanced network (CEEN) for channel matrix estimation.
	
	As illustrated in Fig. \ref{Unet_CE}, the CEEN network leverages the U-Net structure for efficient MIMO channel estimation. It is mainly composed of the encoder, decoder, and refined skip connections.
	Specifically, the encoder module is composed of two units, including a 3 $\times$ 3 kernel convolution layer and a maxpooling layer to extract multilevel discriminative feature maps and generate low-resolution feature maps, while the decoder consists of several convolution layers, concatenation layers, and upsample layers.
	The refined skip connection mainly includes the enhanced depthwise separable convolution (EDSC) module, which enables the decoder to flexibly fuse multiscale features with diverse information, thereby facilitating the effective capture of features.
	For the EDSC module, the input feature maps are first convoluted by a 1 $\times$ 1 convolution layer, and then, the depthwise convolution operation is performed for each input channel. Finally, the output feature maps from the depthwise convolution are convoluted through a 3 $\times$ 3 convolution layer to generate the final feature maps.
	On the decoder side, the data are upsampled and concatenated with the refined features from the EDSC module, enabling a progressive recovery of the detailed and spatial information of the channel feature maps, and ultimately producing the enhanced channel estimated matrix output.

	To ensure high stability under varying wireless MIMO channels, the CEEN is pretrained as an independent module and subsequently embedded into the proposed SISC framework. The raw training data for $\Theta(\cdot)$ are derived from a large dataset $\mathcal H$ consisting of abundant MIMO CSI samples. At each training step, a batch $\mathbf{B} \in \mathcal{H}$ is first sampled.
	Then, the channel estimator $\Theta(\cdot)$ is trained using ground-truth channel $\mathbf{B}$ as labeled training data.
	We adopt MSE as the CEEN loss function $\mathcal{L}_H$, which can be expressed as 
	\begin{equation}
		\mathcal{L}_H = \frac{1}{N^\prime}\sum\limits_{i^\prime=1}^{N^\prime}\|\hat{\mathbf{H}}_{s}(i^\prime) - {\mathbf{B}}(i^\prime)\|^2,
		\label{H_opt}
	\end{equation}
	where $N^\prime$ denotes the number of elements in MIMO fading channel matrix. $\hat{\mathbf{H}}_{s}(i^\prime)$ and ${\mathbf{B}}(i^\prime)$ represent the $i^\prime$-th estimated and ground truth channel matrices, respectively.
	The pretraining method of CEEN is summarized in Algorithm \ref{Alg_H}.

	\begin{figure*}[t]
	\centering
	\setlength{\abovecaptionskip}{-0.1cm} 
	\setlength{\belowcaptionskip}{-2.84em} 
	\includegraphics[width=0.93\linewidth]{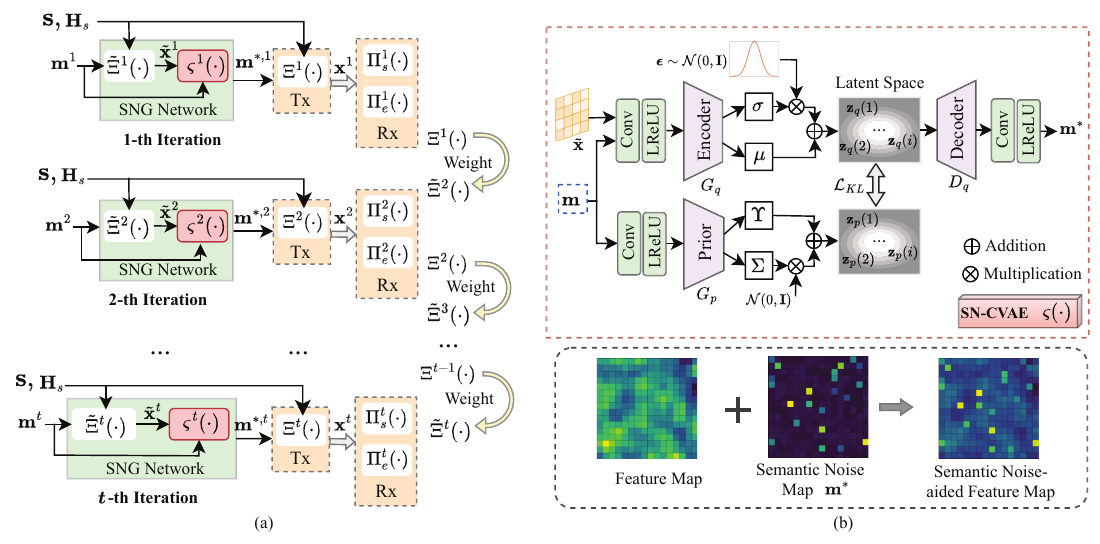}
	\caption{(a) The semantic noise generation mechanism of the SNG network. (b) The detailed SN-CVAE structure.}
	\label{Learnable_SN}
\end{figure*}

	\subsection{Other Network Structure of the SISC Framework}
	The SISC decoder is developed based on the Swin Transformer architecture, where the patch merging blocks are replaced with patch reverse-merging blocks to achieve up-sampling. 
	In the SISC framework, the Channel ModNet pair \cite{ModNet} is incorporated to perform channel coding. This module facilitates the modulation and demodulation of processed semantic features, and generates adaptive codewords to enhance robustness under varying channel conditions.
	
	\section{Learnable Semantic Noise Design}
	\label{sec:SN}
	In this section, we introduce a learnable semantic noise generation mechanism to improve the security of the system to diverse source and channel distributions.

	\subsection{Preprocessing SNG Network} 
	To enhance the system security under varying source semantics and channel conditions, we design a learnable semantic noise generation (SNG) network that leverages an iterative mechanism \cite{YJ_Du}\cite{BY_Xie}.
	The SNG network jointly exploits the source semantic features and the CSI of the legitimate SU to produce adaptive perturbations that are tailored to both content characteristics and channel states. The generated semantic noise is incorporated at the semantic feature level prior to channel transmission, and therefore does not require any additional transmit power.
In the proposed SISC framework, the semantic encoder first compresses raw images into compact and informative representations. The SNG module then acts as the decisive security-enhancing component by reshaping these representations in a content-aware and channel-aware manner. This design enables the legitimate SU to accurately reconstruct the original image, whereas the eavesdropper is unable to do so due to the interference introduced by the semantic noise.
	Furthermore, by adopting learnable semantic noise rather than a fixed one, the proposed system achieves an adaptive trade-off between semantic fidelity and transmission security, mitigating over-distortion for the SU and insufficient perturbation against Eve.

	In Fig. \ref{Learnable_SN}{\red{(a)}}, the SNG network can be viewed as a preprocessing module to generate a suitable $\mathbf{m}$ noise map from a set of predefined semantic noise patterns. 
	For clarity, we denote the whole transmitter of the SISC system as 
	$\Xi(\cdot) = {E_{\varrho}(E_{\theta}(\cdot))}$
	and the receiver as $\Pi(\cdot) = {E_{\varphi}(E_{\phi}(\cdot))}$. To ensure fairness, the decoder of the eavesdropper $\Pi_e(\cdot)$ shares the same network parameters as the legitimate semantic user decoder $\Pi_s(\cdot)$. 
	In each iteration $t$, we aim to obtain an optimal semantic noise map $\mathbf{m}^{*}$ to secure the transmission. 
	However, since the currently transmitted data $\mathbf x^t$ is unavailable for directly guiding the SNG network to optimize $\mathbf{m}$, a proxy encoder $\tilde \Xi^t(\cdot)$ is introduced to generate the codewords $\tilde{\mathbf{x}}^t \in \mathbb C^k$, which can be viewed as the proximate substitute of the actual transmitted codewords ${\mathbf{x}^t}$, i.e.,
	\begin{equation}
		\tilde{\mathbf{x}}^{t}= \tilde \Xi^{t}(\mathbf{s}, \mathbf{H}_s, \mathbf m^{t})=\Xi^{t-1}(\mathbf{s}, \mathbf{H}_s, \mathbf m^{t}),
	\end{equation}
	where the weights of the proxy encoder $\tilde \Xi^t(\cdot)$ are obtained from the latest trained encoder $\Xi^{t-1}(\cdot)$, and the corresponding mean and covariance are also obtained. Then, the predefined stable $\mathbf m$ and the proxy codewords $\tilde{\mathbf{x}}$ after the last iteration are fed to the SN-CVAE \cite{CVAE} for the optimal learnable $\mathbf{m}^*$, which can be expressed as
	\begin{equation}
		\mathbf{m}^* = \varsigma({\mathbf{m}},\tilde{\mathbf{x}}), 
	\end{equation}
	where $\varsigma(\cdot)$ is the proposed SN-CVAE network.

	\subsection{Detailed SN-CVAE Design}
	Based on the aforementioned preprocessing stage, we next present the detailed SN-CVAE design that enables learnable noise generation. The proper semantic noise map at the current stage is unknown, which means that no standard semantic noise values exist for us to evaluate the learnable $\mathbf{m}^*$ through direct criteria such as MSE loss. In this way, the variational learning aided by the predefined condition $\mathbf{m}$, i.e., the generated noise in the previous iteration, is a competitive method for generating $\mathbf{m}^*$ through the latent representation to optimize equation \eqref{p: initial}.

	We propose the SN-CVAE network to generate current optimal proxy recurrent semantic codewords $\tilde{\mathbf{x}}$ with the condition $\mathbf m$. Accordingly, based on the powerful CVAE, which models the latent variables conditioned on side information, the conditional log-likelihood is formulated as 
	\begin{equation}
		\begin{aligned}
			\log p(\tilde{\mathbf{x}}|\mathbf{m})
			&= \log \int p(\tilde{\mathbf{x}},\mathbf{z}|\mathbf{m})\,\mathrm{d}\mathbf{z}\\
			&= \log \int p(\tilde{\mathbf{x}}|\mathbf{z},\mathbf{m})\,p(\mathbf{z}|\mathbf{m})\,\mathrm{d}\mathbf{z},
		\end{aligned}
		\label{log_like}
	\end{equation}
	where $\mathbf{z}$ denotes the latent variable that captures the hidden representation of semantic information conditioned on $\mathbf{m}$. 
	We note that the conditional likelihood $p(\tilde{\mathbf{x}}|\mathbf m)$ involves an intractable marginalization over the latent variable $\mathbf{z}$, i.e., 
	\begin{equation}
		\log p(\tilde{\mathbf{x}}|\mathbf{m})= \log \int p(\mathbf{z}| \tilde{\mathbf{x}}, \mathbf m)\,
		\frac{p(\tilde{\mathbf{x}}, \mathbf{z}|\mathbf m)}{p(\mathbf{z}| \tilde{\mathbf{x}}, \mathbf m)}\ \mathrm{d}\mathbf{z},
	\end{equation}
	where $p(\mathbf{z}| \tilde{\mathbf{x}}, \mathbf m)$ denotes the true posterior distribution of the latent variable.
	In order to infer the true posterior $p(\mathbf{z}| \tilde{\mathbf{x}}, \mathbf m)$, which is generally intractable, we resort to using a variational distribution $q(\mathbf{z}| \tilde{\mathbf{x}}, \mathbf m)$ to approximate it.
	By modifying the evidence lower bound (ELBO) \cite{ELBO}, the objective of SN-CVAE is to maximize the variational lower bound of
	\begin{equation}
		\begin{aligned}
			\label{ELBO}
			\mathcal{L}_\mathrm{SN-CVAE} 
			=& - D_{\mathrm{KL}} \left[ q(\mathbf{z}|\tilde{\mathbf{x}},\mathbf{m}) \big|\big| p(\mathbf{z}|\mathbf{m}) \right] \\
			& + \mathbb{E}_{ q(\mathbf{z}|\tilde{\mathbf{x}},\mathbf{m})} \left[ \log p(\tilde{\mathbf{x}}|\mathbf{z},\mathbf{m}) \right],
		\end{aligned}
	\end{equation}
	where $D_{\mathrm{KL}}[q(\mathbf{z}|\tilde{\mathbf{x}},\mathbf m)||p(\mathbf{z}|\tilde{\mathbf{x}},\mathbf m)]$ is the Kullback-Leibler (KL) divergence \cite{KL} that measures the discrepancy between the true and approximate posterior distributions. Here, $ p(\tilde{\mathbf{x}}|\mathbf{z},\mathbf{m})$ and
	$q(\mathbf{z}|\tilde{\mathbf{x}},\mathbf{m})$ are assumed to be Gaussian distributions. It is noted that the objective of SN-CVAE in Eq. \eqref{ELBO} is to generate task-aware semantic noise that enhances physical-layer security. While the overall system objective in Eq. \eqref{p: initial} aims to minimize the legitimate SU distortion $\mathcal{L}_s$ and maximize the eavesdropper distortion $\mathcal{L}_e$, directly optimizing the noise map $\mathbf {m}$ with respect to $\mathcal{L}_e$ is intractable due to the unknown conditional distribution of optimal perturbations. Therefore, SN-CVAE is introduced as a generative model to learn the conditional distribution, enabling adaptive semantic noise generation conditioned on both source semantics and channel state information.
	
	The detailed SN-CVAE backbone, as shown in Fig. \ref{Learnable_SN}{\red{(b)}}, mainly consists of an encoder, a prior network, and a decoder. 
	Conditioned on the semantic noise $\mathbf m$, the encoder $G_q$ learns the latent distribution $\mathcal N (\mu(\tilde{\mathbf{x}},\mathbf m), \sigma(\tilde{\mathbf{x}},\mathbf m))$ that captures the latent representation of the corresponding semantic codewords $\tilde{\mathbf{x}}$, where $\mu(\tilde{\mathbf{x}},\mathbf m)$ and $\sigma(\tilde{\mathbf{x}},\mathbf m)$ denote the mean and variance of the learned Gaussian model, respectively.
	To ensure that the latent variable $\mathbf z$ sampled during inference remains semantically consistent with the input ${\mathbf{m}}$, a prior network $G_p$ is introduced to enforce consistency between the latent distributions obtained in training and inference. 
	The prior network encodes a given semantic noise map into a conditional prior distribution $\mathcal N (\Upsilon(\mathbf m), \Sigma(\mathbf m))$ with mean $\Upsilon(\mathbf m)$ and variance $\Sigma(\mathbf m) = diag(\Sigma_1^2, \Sigma_2^2, \cdots, \Sigma_n^2)$, which models the underlying variability of semantic noise. Conditioned on the sampled latent variable $\mathbf z$ and the semantic codewords $\tilde{\mathbf{x}}$, the decoder $D_q$ subsequently generates the optimized semantic noise map $\mathbf m^*$.
	To compute the gradient, we use reparameterization techniques \cite{reparameter} to sample the latent variable $\mathbf{z}$ via 
	\begin{equation}
		\mathbf{z}=\mu(\tilde{\mathbf{x}},\mathbf m)+\boldsymbol{\epsilon}\odot \sigma(\tilde{\mathbf{x}},\mathbf m),
	\end{equation}
	where $\boldsymbol{\epsilon} \sim \mathcal{N}(0, \mathbf I)$ is the sampled Gaussian distribution and $\odot$ denotes the Hadamard product. 
	
	To learn the CVAE backbone network, we need to maximize
	the conditional variational lower bound defined in Eq. \eqref{ELBO}. 
	The first term in Eq. \eqref{ELBO} acts as a regularization term to minimize the difference between the data-conditional distribution
	$q(\mathbf{z}|\tilde{\mathbf{x}},\mathbf{m})$ and the prior distribution $p(\mathbf{z}|\mathbf{m})$. 
	Here, we take KL divergence as the penalty function to
	minimize the gap between the two Gaussian distributions
	$q(\mathbf{z}|\tilde{\mathbf{x}},\mathbf{m})$ and $p(\mathbf{z}|\mathbf{m})$.
	The second term in Eq. \eqref{ELBO} is the
	reconstruction error measuring the information loss between the generated data $\bar{\mathbf{x}}$ and the original data $\tilde{\mathbf{x}}$ with condition $\mathbf{m}$. We maximize the conditional log-likelihood $\mathbb{E}_{ q(\mathbf{z}|\tilde{\mathbf{x}},\mathbf{m})} \left[ \log p(\tilde{\mathbf{x}}|\mathbf{z},\mathbf{m}) \right]$ for accurate reconstruction. In practice, the loss can be
	computed as the MSE distortion between $\tilde{\mathbf{x}}$ and $\bar{\mathbf{x}}$.
	Finally, with the trained latent variable $\mathbf{z}$ from the encoder ${G}_q(\cdot)$, we utilize the decoder $D_q(\cdot)$ to obtain the optimal $\mathbf m^*$.
	
	\subsection{Loss for SN-CVAE}
	The SN-CVAE network is trained to maximize
	the conditional log-likelihood of the first term in Eq. \eqref{ELBO}.
	Since this objective function is intractable, we instead maximize its variational lower bound. We minimize
	the KL divergence between the data-conditional distribution
	$q(\mathbf{z}|\tilde{\mathbf{x}},\mathbf{m})$ and the prior distribution $p(\mathbf{z}|\mathbf{m})$, to mitigate the
	distortion between the encoding of latent variables at
	learning and inference stages:
	\begin{equation}
		\begin{aligned}
			\mathcal{L}_{\mathrm{KL}}=\sum_{i=1}^{N}q(\mathbf{z}(i)|\tilde{\mathbf{x}}(i),\mathbf{m}(i))\log\left(\frac{q(\mathbf{z}(i)|\tilde{\mathbf{x}}(i),\mathbf{m}(i))}{p(\mathbf{z}(i)|\tilde{\mathbf{x}}(i))}\right),
		\end{aligned}
		\vspace{-0.1cm}
	\end{equation}
	where $q(\mathbf{z}(i)|\tilde{\mathbf{x}}(i),\mathbf{m}(i)) = \mathcal N (\mu(\tilde{\mathbf{x}},\mathbf m), \sigma(\tilde{\mathbf{x}},\mathbf m))$, $p(\mathbf{z}(i)|\tilde{\mathbf{x}}(i)) = \mathcal N (\Upsilon(\mathbf m), \Sigma(\mathbf m))$ and $N$ is the number of training images. 
	To maximize $\mathbb{E}_{ q(\mathbf{z}|\tilde{\mathbf{x}},\mathbf{m})} \left[ \log p(\tilde{\mathbf{x}}|\mathbf{z},\mathbf{m}) \right] $ for the reconstruction of $\tilde{\mathbf{x}}$, we define the loss $L_\mathrm{rec}$ as follows
	\begin{equation}
		\mathcal L_{\mathrm{rec}}=\frac{1}{N}\sum_{i=1}^{N}||\tilde{\mathbf{x}}(i)-\bar{\mathbf{x}}(i)||^{2}. 
	\end{equation}
	In summary, the loss $\mathcal{L}_\mathrm{SN-CVAE}$ is defined as the sum of $\mathcal{L}_{\mathrm{KL}}$ and $\mathcal L_{\mathrm{rec}}$
	\begin{equation}
		\mathcal{L}_\mathrm{SN-CVAE} =\mathcal L_{\mathrm{rec}} + \beta \mathcal L_{\mathrm{KL}},
	\end{equation}
	where $\beta > 0$ is a regularization parameter.
	Accordingly, given the optimized transceiver beamformer $\mathbf P^*$ and $\mathbf U_s^*$, the overall training loss for the SISC framework is defined as the combination of the deep JSCC loss and the ELBO loss from the SN-CVAE, which can be formulated as
	\begin{equation}
		\mathcal{L}^\prime =\mathcal{L}(\mathbf m) + \eta \mathcal{L}_\mathrm{SN-CVAE},
	\end{equation}
	where $\eta$ is the trade-off parameter.
	Based on the different source images and SU channel conditions, the semantic noise-aided communication framework can effectively weaken the data reconstruction performance of eavesdroppers.

		\begin{algorithm}[tbp]
		\caption{Transceiver Beamformer Design}
		\label{alg: SCA}
		\textbf{Input:} SU channel matrix $\mathbf{H}_s$, noise covariance $\sigma_s^2$, maximum transmit power $P_\mathrm{max}$, delay threshold $T_\mathrm{max}$. \\
		\textbf{Output:} Optimized $\mathbf{P}^*$ and $\mathbf{U}_s^*$.
		
		\begin{algorithmic}[1]
			\STATE \textbf{Initialize:} 
			Transceiver beamformer $\mathbf{P}^{(0)}$ and $\mathbf{U}_s^{(0)}$, accumulation vectors $\Phi_{\mathbf{P}}^{(0)} = \Phi_{\mathbf{U}_s}^{(0)} = 0$, $\mathbf{J}_{\mathbf{P}}^{(0)} \!=\! 0$, $J_s^{(0)} \!=\! 0$, and iteration index $n \!=\! 0$.
			\FOR{$n=1, 2, ...$}
			\STATE Construct the surrogate functions w.r.t.
			$\mathbf{P}$ and $\mathbf{U}_s$ according to Eq. \eqref{surrogate}.
			\STATE Solve the surrogate problem in \eqref{p:surrogate}
			\STATE Update transceiver beamformer$\mathbf{P}^{(n+1)}$ and $\mathbf{U}_s^{(n+1)}$ according to Eq. \eqref{eq: transmit update}.
			\ENDFOR
			\STATE Until the convergence criteria is satisfied.
		\end{algorithmic}
	\end{algorithm}

	\section{Transceiver Beamformer Optimization Design}
	\label{sec:Transceiver}
	
	In this section, given the optimized semantic noise map $\mathbf m^*$ obtained in the previous stage, we focus on the joint transceiver beamforming optimization to enhance the security and efficiency of semantic image transmission. Then, a low-complexity optimization algorithm is developed to design the transceiver beamformer efficiently.

	\subsection{Transceiver Beamformer Problem Design} 
	After obtaining the optimized semantic noise map $\mathbf m^*$, the transceiver beamformer design objective is accordingly transformed into 
	\begin{equation}
		\mathcal{L}(\mathbf{P}, \mathbf{U}_s, \mathbf m^*) \triangleq \mathcal{L}_s - \lambda \mathcal{L}_e,
		\label{p: v}
	\end{equation}
	given the power constraint in Eq. \eqref{power_con} and the transmission delay constraint \eqref{delay}. Thus, the transceiver beamformer optimization problem can be formulated as
	\begin{subequations}
		\label{Problem}
		\begin{align}
			\min\limits_{\mathbf P, \mathbf U_s}\quad
			&\mathcal{L}(\mathbf{P}, \mathbf{U}_s) \triangleq \mathcal{L}_s - \lambda \mathcal{L}_e \label{a}\\
			\mathrm{s.t.}\,\,\,\,
			& \mathrm{Tr}(\mathbf{P}\mathbf{P}^H) \leq P_\mathrm{max} \label{b}, \\
			&T_s \leq T_\mathrm {max} \label{c},
		\end{align}
	\end{subequations}
	where \eqref{b} denotes the transmit power constraint at the BS, and \eqref{c} specifies the maximum tolerable transmission delay of the SU, denoted by $T_\mathrm {max}$. 
	
	\subsection{Optimization Algorithm Design}
	We propose to leverage the constrained successive convex approximation (CSSCA) \cite{CSSCA} technique to update the transceiver beamformer. Due to the non-convexity of the objective function and the coupling between $\mathbf{P}$ and $\mathbf{U}_s$, the global optimum is intractable. Thus, the algorithm converges to a stationary point, yielding an efficient suboptimal solution. To facilitate the application of CSSCA, we first construct a surrogate function satisfying the gradient consistency for the objective function \eqref{a} and the upper bound property for the delay constraint \eqref{c}.
	Specifically, at iteration $n$, we adopt the first-order expansion to approximate \eqref{a}, and the surrogate objective function constructed at $\mathbf P = \mathbf P^{(n)}$ and $\mathbf U_s = \mathbf U_s^{(n)}$ is shown as follows
	\vspace{0.1cm}
	\begin{equation}
		\label{surrogate}
		\begin{aligned}
			&\mathcal{\tilde{L}}(\mathbf P, \mathbf U_s, \mathbf P^{(n)}, \mathbf U_s^{(n)}) \\
			&\triangleq 2\langle \tau^{(n)} \nabla_{\mathbf P} \mathcal L^{(n)} + (1-\tau^{(n)}) \Phi_{\mathbf P}^{(n)}, \mathbf P - \mathbf P^{(n)}\rangle \\
			&\quad+ 2\langle \tau^{(n)} \nabla_{\mathbf U_s} \mathcal L^{(n)} + (1-\tau^{(n)}) \Phi_{\mathbf U_s}^{(n)}, \mathbf U_s - \mathbf U_s^{(n)}\rangle\\
			&\quad+C_{\mathbf P} \|\mathbf P-\mathbf P^{(n)}\|^2 + C_{\mathbf U_s} \|\mathbf U_s-\mathbf U_s^{(n)}\|^2,
		\end{aligned}
		\vspace{0.1cm}
	\end{equation}
	where $C_{\mathbf P} > 0$ and $C_{\mathbf U_s} > 0$ are constants to ensure strong convexity. $\Phi_{\mathbf P}^{(n)}$ and $\Phi_{\mathbf U_s}^{(n)}$ are the accumulation vectors that can be updated recursively as
	\begin{equation}
		\begin{aligned}
			\Phi_{\mathbf P}^{(n)} &= \tau^{(n)} \nabla_{\mathbf P} \mathcal L^{(n)} + (1-\tau^{(n)}) \Phi_{\mathbf P}^{(n-1)},\\
			\Phi_{\mathbf U_s}^{(n)} &= \tau^{(n)} \nabla_{\mathbf U_s} \mathcal L^{(n)} + (1-\tau^{(n)}) \Phi_{\mathbf U_s}^{(n-1)},
		\end{aligned}
	\end{equation}
	with $\tau^{(n)}$ being a sequence to be properly chosen \cite[Assumption 5]{CSSCA}, which satisfies 
	\begin{equation}
		\begin{aligned}
			&\lim_{n\to\infty}\tau^{(n)}=0,\\
			&\sum\limits_{n=0}^{\infty} \tau^{(n)}= \infty,\\
			&\sum\limits_{n=0}^{\infty} (\tau^{(n)})^2 < \infty.
		\end{aligned}
		\label{gamma_con}
	\end{equation}
Here, the gradients $\nabla_{P}\mathcal{L}^{(n)}$ and $\nabla_{U_s}\mathcal{L}^{(n)}$ capture the sensitivity of the semantic loss with respect to the transceiver beamformers.
	We consider a linear receive combiner at the SU receiver side, and the achievable rate can be approximated as
	\vspace{0.1cm}
	\begin{equation}
		\log \det\left(1+\frac{1}{\sigma_s^2}\|\mathbf{U}_s^H\mathbf{H}_s\mathbf{P}\|_F^2\right) \geq \frac{k}{BT_\mathrm{max}}.
	\end{equation} 
	\vspace{0.1cm}
	In this way, the constraint in \eqref{c} is equivalent to
	\vspace{0.1cm}
	\begin{equation}
		\begin{aligned}
			&J_s(\mathbf P, \mathbf {U}_{s}) \triangleq \xi_s^\mathrm {min} \sigma_{s}^2 -\|\mathbf{U}_{s}^H\mathbf{H}_{s}\mathbf{P}\|^2_F \leq 0,
		\end{aligned}
		\vspace{0.1cm}
	\end{equation}
	where $\xi_s^\mathrm {min} \triangleq 2^{k/BT_\mathrm{max}} - 1$. 
	To facilitate the iterative optimization of the beamformers, at iteration $n$,
	$J_s(\mathbf P, \mathbf {U}_{s})$ can be approximated by 
	\vspace{0.1cm}
	\begin{equation}
		\label{eq perfect: stationary point snr}
		\begin{aligned}
			&\tilde{J}_s(\mathbf P, \mathbf P^{(n)}, \mathbf {U}_{s}, \mathbf {U}_{s}^{(n)}) \\
			&\triangleq (1-\tau^{(n)})J_s^{(n-1)} -\tau^{(n)}\|\mathbf{U}_{s}^H\mathbf{H}_{s}\mathbf{P}\|^2_F \\
			& +\tau^{(n)}\xi_s^\mathrm {min} \sigma_{s}^2\|\mathbf{U}_{s}\|^2_F -2\langle \mathbf P - \mathbf P^{(n)}, \tau^{(n)}\mathbf H_{s} \mathbf H_{s}^H \mathbf P^{(n)}\\
			&+ (1-\tau^{(n)}) \mathbf J_{\mathbf P}^{(n)} \rangle +C_{\mathbf P} \|\mathbf P-\mathbf P^{(n)}\|^2 + C_{\mathbf {U}_{s}} \|\mathbf {U}_{s} - \mathbf {U}_{s}^{(n)}\|^2.
		\end{aligned}
		\vspace{0.1cm}
	\end{equation}
	Here, both $C_{\mathbf P} > 0$ and $C_{\mathbf {U}_{s}} > 0$ are constants to ensure strong convexity of the surrogate function.
	$\mathbf J_{\mathbf P}^{(n)}$ and $J_s^{(n)}$ are the accumulation functions that can be updated recursively as follows
	\begin{equation}
		\begin{aligned}
			\mathbf J_{\mathbf P}^{(n)} &= \tau^{(n)}\mathbf H_{s} \mathbf H_{s}^H \mathbf P^{(n)} + (1-\tau^{(n)}) \mathbf J_{\mathbf P}^{(n-1)},\\
			J_s^{(n)} &= \tau^{(n)} J_s(\mathbf P^{(n)}, \mathbf U^{(n)}) + (1-\tau^{(n)}) J_s^{(n-1)}.
		\end{aligned}
	\end{equation}
	Given $(\mathbf P^{(n)}, \mathbf {U}_{s}^{(n)})$, we can solve the following surrogate problem 
	\begin{subequations}
		\label{p:surrogate}
		\begin{align}
			(\hat{\mathbf P}^{(n)}, \hat{\mathbf U}_s^{(n)})
			&\triangleq\arg\min\limits_{\mathbf P, \mathbf {U}_{s}}\quad
			\tilde{\mathcal L}(\mathbf P, \mathbf {U}_{s}, \mathbf P^{(n)}, \mathbf {U}_{s}^{(n)}) \\
			\mathrm{s.t.}\,\,\,\,
			&\mathrm{Tr}(\mathbf{P}\mathbf{P}^H) \leq P_\mathrm{max}, \\
			& \tilde{J}_s(\mathbf P, \mathbf P^{(n)}, \mathbf {U}_{s}, \mathbf {U}_{s}^{(n)}) \leq 0.
		\end{align}
	\end{subequations}
	We observe that the problem is a convex quadratic programming (QP) problem, which can be effectively solved by CVXPY. The transceiver beamformer $\mathbf P^{(n+1)}$ and $\mathbf {U}_{s}^{(n+1)}$ can be respectively updated by
	\begin{equation}
		\label{eq: transmit update}
		\begin{aligned}
			&\mathbf P^{(n+1)} = \mathbf P^{(n)} + \epsilon^{(n)} (\hat{\mathbf{P}}^{(n)} - \mathbf P^{(n)}),\\
			&\mathbf U_s^{(n+1)} = \mathbf U_s^{(n)} + \epsilon^{(n)} (\hat{\mathbf U_s}^{(n)} - \mathbf U_s^{(n)}),
		\end{aligned}
	\end{equation}
	where $ \epsilon^{(n)}$ is the iterative step size for updating $\mathbf P$ and $\mathbf U_s$ \cite[Assumption 5]{CSSCA}: $\lim_{n\to\infty} \epsilon^{(n)}/\tau^{(n)}=0$. The above procedure is summarized in Algorithm \ref{alg: SCA}.

	\subsection{Convergence and Complexity Analysis}
	\emph{Convergence analysis:} Based on the constrained optimization algorithm for transceiver beamformer design, a stationary point is guaranteed for this problem. Therefore, the initial equivalent function is guaranteed to decrease monotonically, and our proposed algorithm is convergent.
	
	\emph{Complexity analysis:}
	The complexity of gradient computation and the construction of the surrogate terms is $\mathcal O(N_r N_t k + N_t^2 k)$. Solving the convex QP subproblem in each iteration involves a complexity on the order of $\mathcal O((2N_t k)^3)$. Thus, the overall per-iteration computational complexity can be summarized as $\mathcal O(N_r N_t k + N_t^2 k + (2N_t k)^3)$.

	\subsection{Overall Training and Deployment Strategy}
	Our goal is to obtain suitable SISC parameters
	to enhance the SU image reconstruction quality while interfering with the Eve over the MIMO wiretap channel. To realize it, we first pretrain the CEEN by using sampled channel data under different channel conditions. The pretraining process is summarized in Algorithm \ref{Alg_H}, where the channel state is assumed to remain unchanged during CSI feedback.
	Next, as illustrated in Algorithm \ref{alg: SCA}, the transceiver beamformer is well designed to improve the system throughput and transmission efficiency. For practical deployment, we jointly train the SISC framework together with CEEN and the transceiver beamformer design in an end-to-end manner. By leveraging the pretrained CEEN as an effective initialization, accurate feedback CSI are enabled in the learnable semantic communication framework, which helps the overall SISC framework converge fast and stably.

	\section{Numerical Results}
	\label{sec:Results}
	In this section, numerical simulations are conducted to evaluate the effectiveness of the proposed algorithm for wireless image transmission over wiretap channels.
	
	\subsection{Experimental Setups}
	\subsubsection{Datasets} 
	For wireless secure semantic image transmission, we utilize the Cityscapes dataset \cite{Cityscapes} for training and select the CVRG-Pano dataset \cite{CVRG} as the testing dataset. 
	The Cityscapes dataset comprises street-scene images collected from 50 urban locations under consistent lighting and weather conditions. The CVRG-Pano dataset contains 600 high-resolution panoramic images (1664 $\times$ 832 pixels) with pixel-level labels spanning 20 semantic classes grouped into 7 categories. To enhance sample diversity and model generalization, data augmentation techniques, such as random flipping and random cropping to 256 $\times$ 256, are employed.

		\begin{figure*}[htpb]
		\centering
		\captionsetup[subfloat]{labelsep=none,format=plain,labelformat=empty}
		\subfloat{
			\label{PSNR_SNR}
			\begin{minipage}[t]{0.32\linewidth}
				\includegraphics[width=\textwidth]{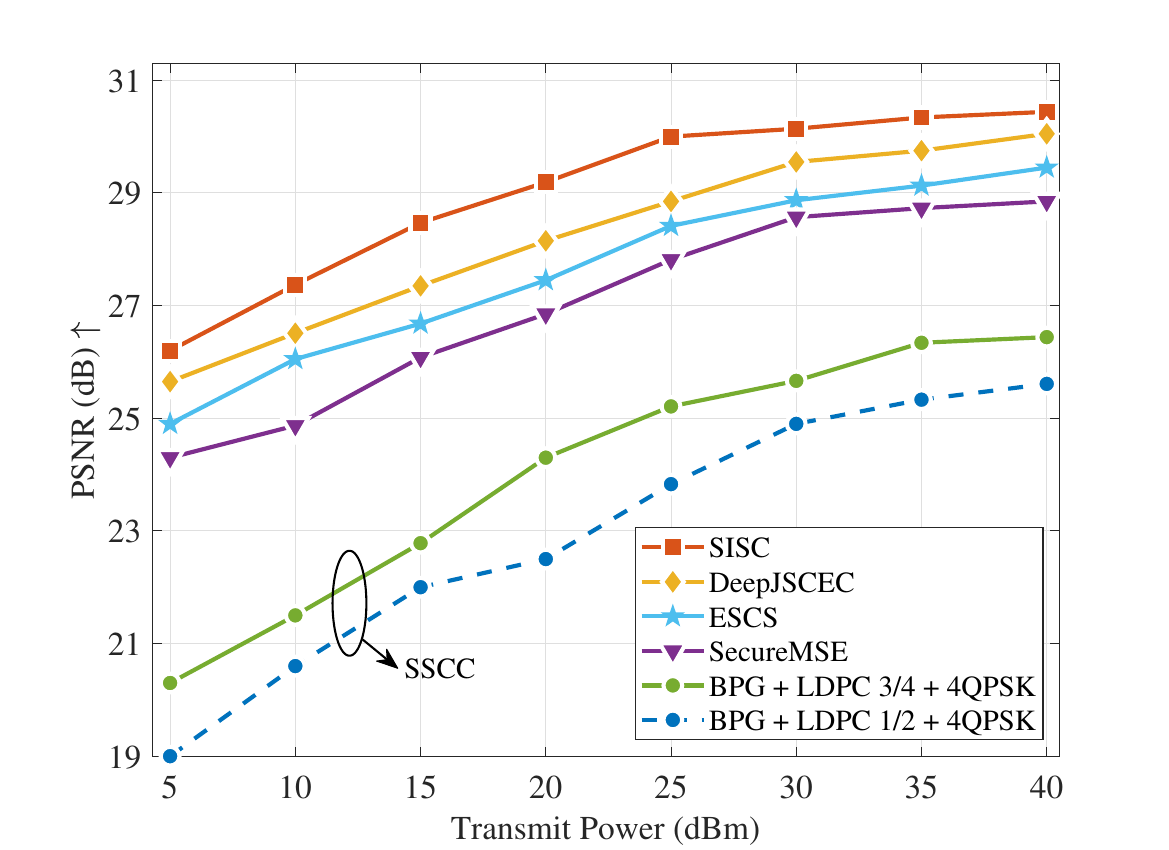}
			\end{minipage}%
		}\hfill
		\subfloat{
			\label{SSIM_SNR}
			\begin{minipage}[t]{0.32\linewidth}
				\includegraphics[width=\textwidth]{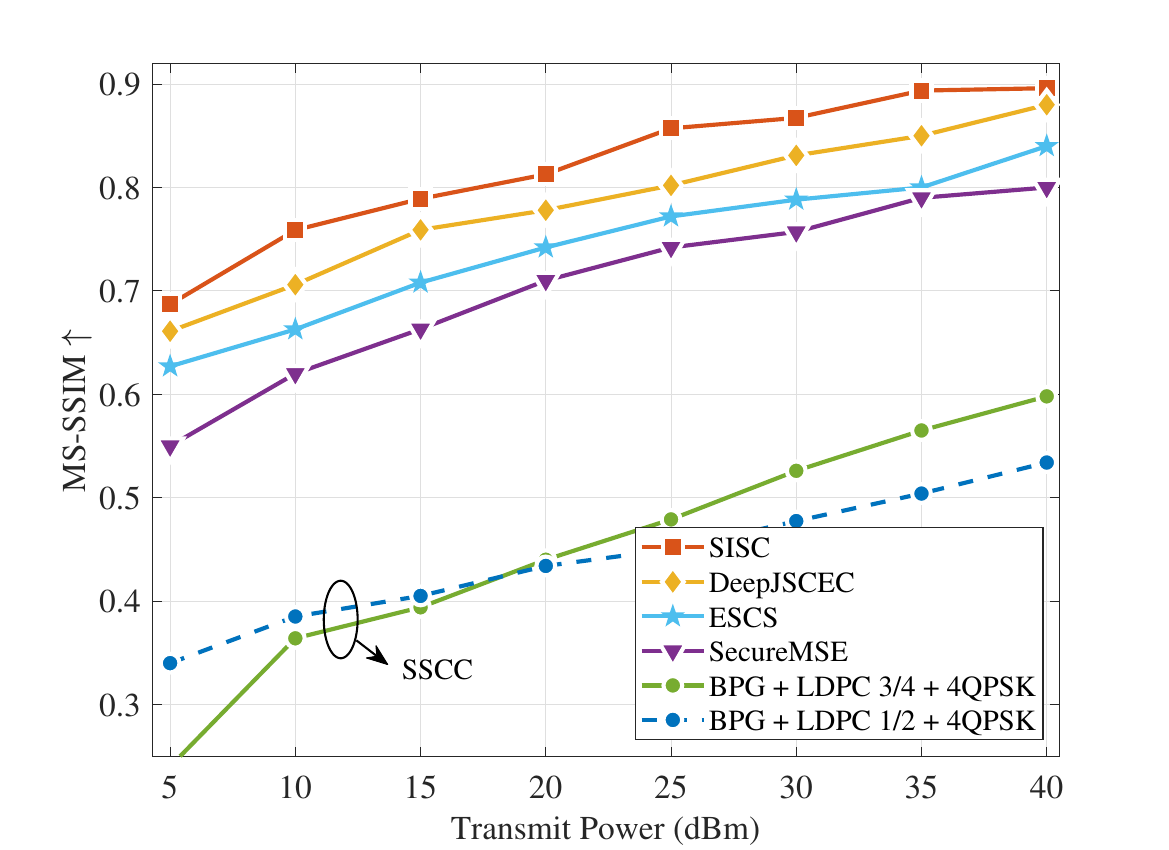}
			\end{minipage}%
		}\hfill
		\subfloat{
			\label{LPIPS_SNR}
			\begin{minipage}[t]{0.32\linewidth}
				\includegraphics[width=\textwidth]{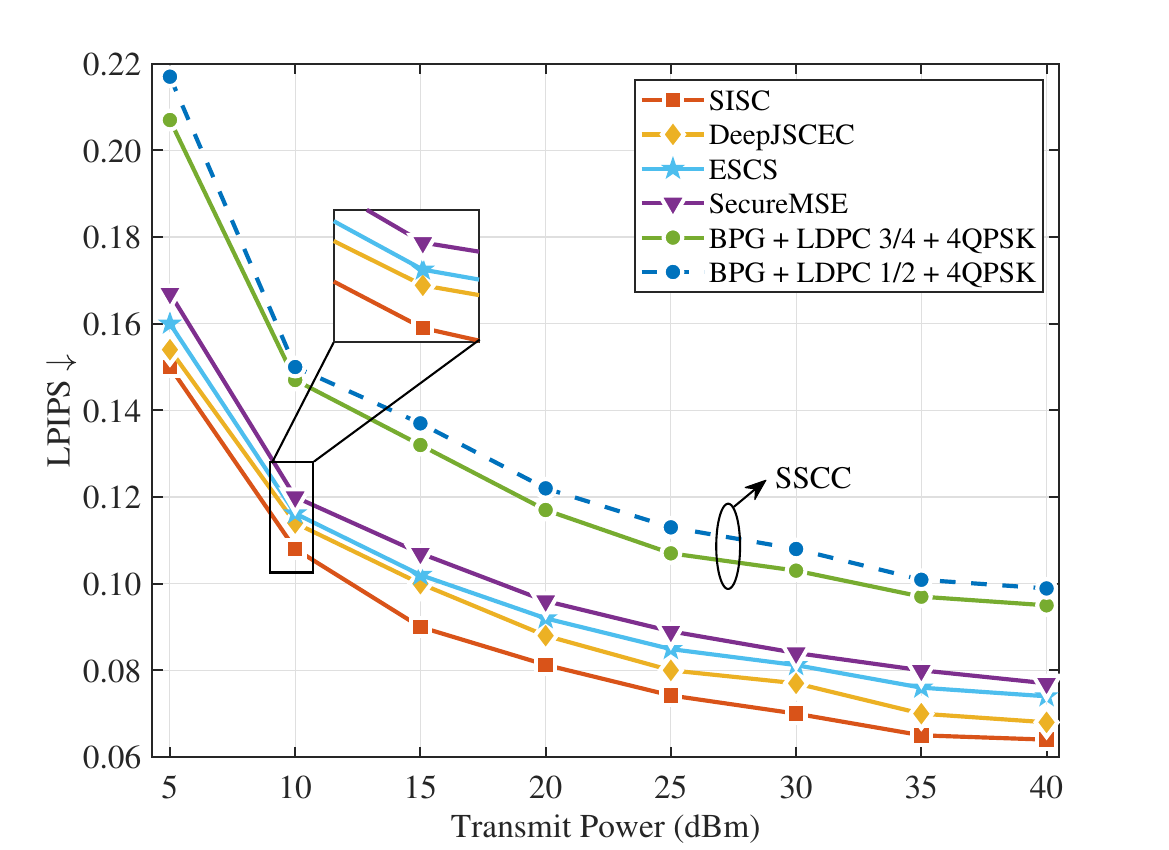}
			\end{minipage}%
		}\hfill
		\caption{Performance comparison of different schemes versus transmit power on the CVRG-Pano dataset over MIMO fading channels.}
		\label{SNR}
		\vspace{-0.2cm}
	\end{figure*}

	\subsubsection{Parameters and Training Details}
	We utilize the Swin Transformer network as the semantic codec backbone with $\{N_1, N_2, N_3, N_4\} = \{2, 2, 6, 2\}$ transformer blocks and the Channel ModNet as the channel codec. The proxy JSCC encoder has the same network deployment as above. In our simulations, a $2 \times 2$ MIMO system is considered for the physical layer secure transmission. In order to ensure proper semantic noise level acquisition, we set
	the range of $\mathbf m$ from $1 \times 10^{-3}$ to $1.5 \times 10^{-2}$, as a smaller semantic noise level tends to retain the essential semantics delivered to SU. 
	The power budget and the maximum tolerable delay are w.l.o.g. set to $P_\mathrm{max}=$ 50 dBm and $T_\mathrm{max}=$ 1 ms, respectively. Total channel bandwidth $B$ is 5 MHz.
	The Adam optimizer \cite{Adam} is adopted for model training, where the variable learning rate decreases step-by-step from $1 \times 10^{-4}$ to $2 \times 10^{-5}$. The batch size is configured as 16.
	\subsubsection{Evaluation Metrics}
	We utilize the pixel-wise metric peak signal-to-noise ratio (PSNR), the perceptual-level multi-scale structural similarity (MS-SSIM), and the learned perceptual image patch similarity (LPIPS) as measurements for the reconstructed image quality. Higher PSNR/MS-SSIM values and lower LPIPS values indicate better performance. These metrics are selected because they jointly characterize both distortion-level and perceptual-level image quality, which is consistent with the design objective of the proposed SISC framework, i.e., preserving high-quality image reconstruction at the legitimate SU while effectively degrading the semantic reconstruction capability of Eve. 
	
	\subsubsection{Comparison Schemes}
	We consider the DL-based transmission method and the traditional separate source-channel coding (SSCC) as the baseline schemes. 

	\begin{figure}[t]
	\centering
	\setlength{\abovecaptionskip}{0.0cm} 
	\setlength{\belowcaptionskip}{-4.84em} 
	\subfloat{
		\includegraphics[width=0.95\linewidth]{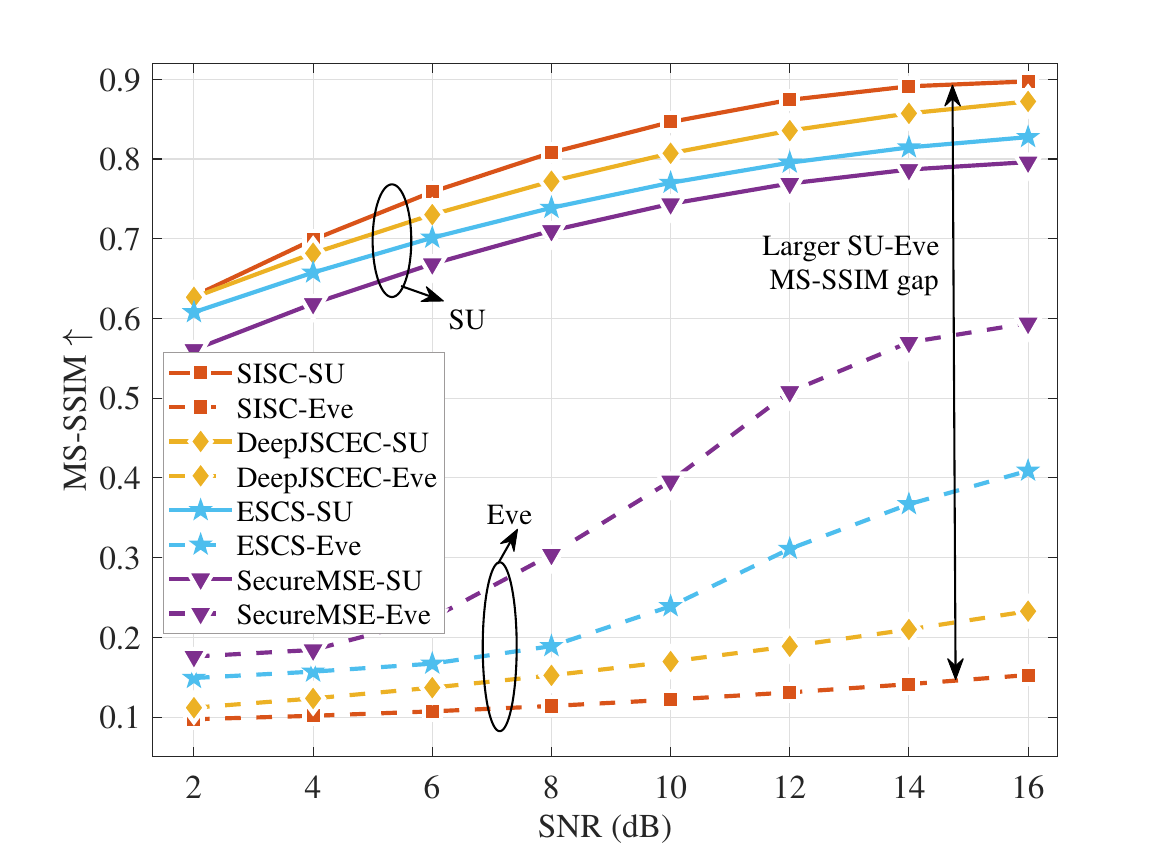}}
	\caption{Security performance comparison of different schemes for SU and Eve in MIMO fading wiretap system.}
	\label{SSIM_SU_Eve}
\end{figure}

	\begin{itemize}		
		\item {\bf{DeepJSCEC}}. For the JSCC end-to-end coding framework, the DeepJSCEC baseline in \cite{DeepJSCEC} provides security against chosen-plaintext attacks from the eavesdropper. 
		
		\item {\bf{SecureMSE}}. The image semantic encryption scheme \cite{Maojun Zhang} employs a residual-structured joint source–channel autoencoder to extract and transmit semantic features. 
				
		\item {\bf{ESCS}}. The scheme in \cite{Xinlai Luo} adopts an adversarial semantic encryption strategy to maintain communication accuracy, which is originally designed for text transmission. For fair comparison, the text source is replaced with an image source, while other components, including the neural network–based encryption/decryption modules and key selection mechanism, remain unchanged.

		\item {SSCC.} The digital baseline employs BPG image codec with low-density parity-check (LDPC) for channel coding, quadrature phase shift keying (QPSK) for modulation, and the singular value decomposition (SVD) for precoding. The coding rates of LDPC are set to be 3/4 and 1/2, denoted as {\bf{BPG+LDPC3/4+4QPSK}} and {\bf{BPG+LDPC1/2+4QPSK}}, respectively.
		
	\end{itemize}

	\subsection{Experimental Results}

				\begin{figure}[t]
		\centering
		\setlength{\abovecaptionskip}{0cm} 
		\setlength{\belowcaptionskip}{-4.84em} 
		\subfloat{
			\includegraphics[width=0.95\linewidth]{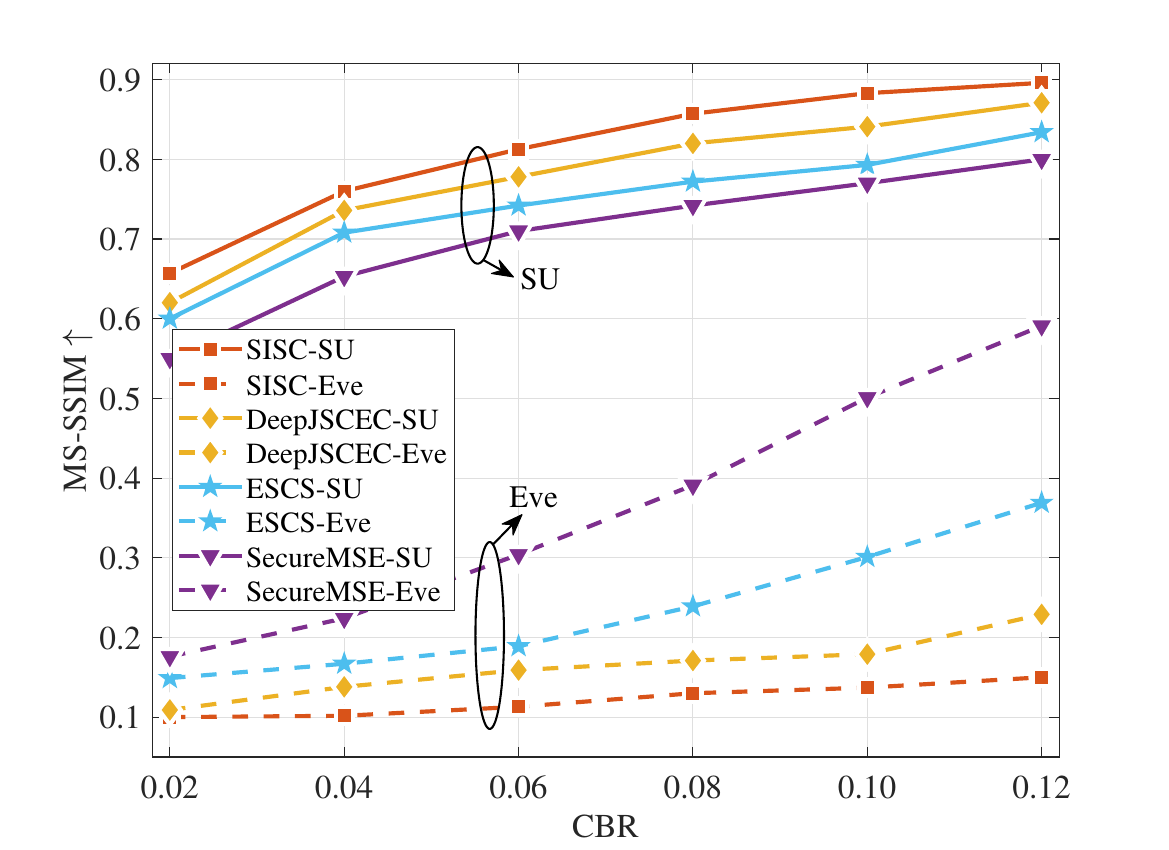}}
		\caption{MS-SSIM performance versus different CBRs in the MIMO fading channel with $P_\mathrm{max}$ = 35 dBm.}
		\label{SSIM_CBR}
	\end{figure}

	\subsubsection{Image Reconstruction Performance}
	Fig. \ref{SNR}\subref{PSNR_SNR} illustrates the PSNR difference between the legitimate SU and the eavesdropper Eve as a function of varying channel conditions with CBR set to $\rho = 1/12$.
	The results indicate that as the transmit power increases, the image recovery performance gradually improves and tends to stabilize after reaching a certain value.
	We observe that the proposed SISC scheme significantly outperforms the other baselines under all power budgets, with the largest gain achieved under $P$ = 10dBm, about 2.35 dB over SecureMSE scheme. Compared to SSCC schemes, SISC provides significantly larger performance gain since traditional schemes would be confronted with serious cliff effects with harsh channel conditions.
	In addition, the MS-SSIM performance of different schemes are shown under different transmission powers in Fig. \ref{SNR}\subref{SSIM_SNR}. The DL-based schemes achieve better reconstruction results compared to the traditional SSCC schemes, which indicates that satisfactory perceptual quality is achieved by effectively exploiting semantic representations embedded in the image data. 
	Compared with the DeepJSCEC, ESCS, and SecureMSE schemes, the proposed SISC achieves an average improvement of 3.68\%, 7.77\%, and 11.62\% in MS-SSIM, respectively. 
As illustrated in Fig. \ref{SNR}\subref{LPIPS_SNR}, the DL-based schemes consistently outperform the traditional separated coding in terms of LPIPS. In the considered system, the lower LPIPS values indicate the better reconstruction quality for the legitimate user, demonstrating that effective semantic feature extraction from the original image signals ensures satisfactory visual perception performance.
	This result demonstrates that SISC not only reconstructs images well but also maintains high visual quality for human perception under MIMO fading channels.

\begin{figure}[t]
	\centering
	\centering
	\setlength{\abovecaptionskip}{0.1cm} 
	\setlength{\belowcaptionskip}{-4.84em} 
	\subfloat{
		\includegraphics[width=0.99\linewidth]{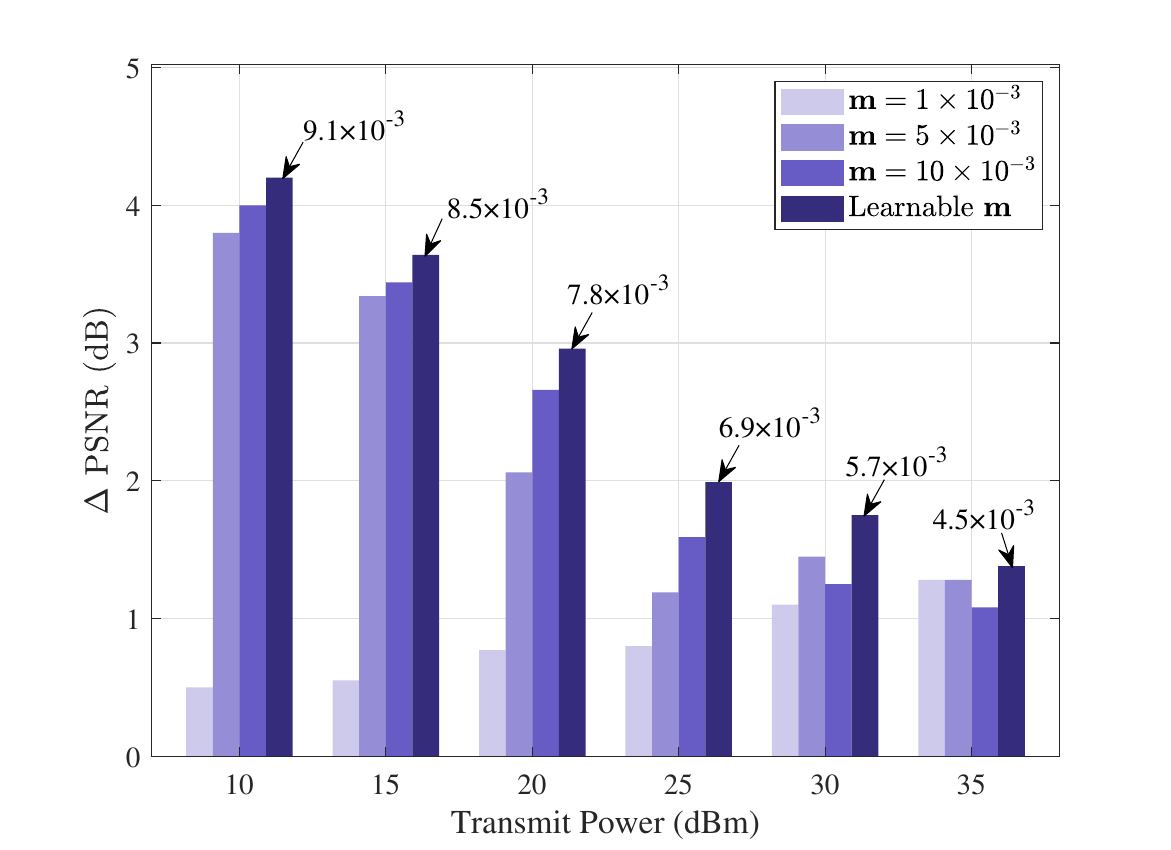}}
	\caption{Comparisons of PSNR improvement of varying learnable semantic noise levels at different transmit powers.}
	\label{Learnable_m}
\end{figure}

\begin{table}[t]
	\centering
	\caption{PSNR Performance Evaluation of SU and Eve Under Different MIMO Modes. Red Marks Denote the Performance Gap Between the Legitimate SU and Eve.}
	\begin{tabular}{c|c|c}
		\hline
		MIMO Type & SU (dB) & Eve (dB) \\
		\hline
		2 $\times$ 2 Antennas & 29.76 {\red{(+19.39)}} & 10.37 \\
		4 $\times$ 4 Antennas & 31.08 {\red{(+20.13)}} & 10.95 \\
		8 $\times$ 8 Antennas & 32.14 {\red{(+20.88)}} & 11.26 \\
		\hline
	\end{tabular}
	\label{MIMO}
	\vspace{-0.24cm}
\end{table}
	
	\subsubsection{Security Evaluation}
The performance comparison between the proposed SISC scheme and the DL-based baselines under the MIMO wiretap system is shown in Fig. \ref{SSIM_SU_Eve}. The channel SNR is uniformly sampled from the range of [2, 16] dB. It can be observed that the proposed SISC scheme consistently achieves the highest reconstruction performance at the SU side, while maintaining a low and nearly constant reconstruction quality at the Eve side, even as the SNR increases. In contrast, the reconstruction quality of Eve in other schemes, particularly the SecureMSE scheme, rises significantly with increasing SNR, indicating a potential risk of information leakage and thus degraded transmission security.
The superiority of the proposed SISC scheme stems from the introduction of a learnable semantic noise map that is generated based on the legitimate user’s source and channel conditions. This semantic noise is optimized with the encoder-decoder pair, enabling SU to adaptively eliminate its effect while making it unpredictable and non-invertible to Eve. Since Eve lacks access to the semantic-noise generation mechanism and its conditioning information, the reconstructed features at Eve remain semantically distorted even at good channel conditions. In contrast, these baseline schemes, which lack semantic-level perturbation, allow Eve’s reconstruction quality to increase rapidly, leading to potential information leakage.

	To further investigate the impact of CBR constraints on the transmission efficiency and security, we evaluate the MS-SSIM performance under different CBR values. As illustrated in Fig. \ref{SSIM_CBR}, the reconstruction quality of the SU generally improves as the CBR increases, since a larger bandwidth ratio allows for the transmission of more semantic features. Furthermore, regarding security, while the increased bandwidincreasedth potentially offers Eve increased information, the proposed SISC scheme successfully limits Eve's reconstruction quality at a low level. This result demonstrates the effectiveness and robustness of the proposed semantic noise design and transmit beamformer algorithm, which incorporates the CSI of the SU to enhance wireless secure image transmission.

\begin{figure}[t]
	\centering
	\setlength{\abovecaptionskip}{0.1cm} 
	\setlength{\belowcaptionskip}{-4.84em} 
	\subfloat{
		\includegraphics[width=0.99\linewidth]{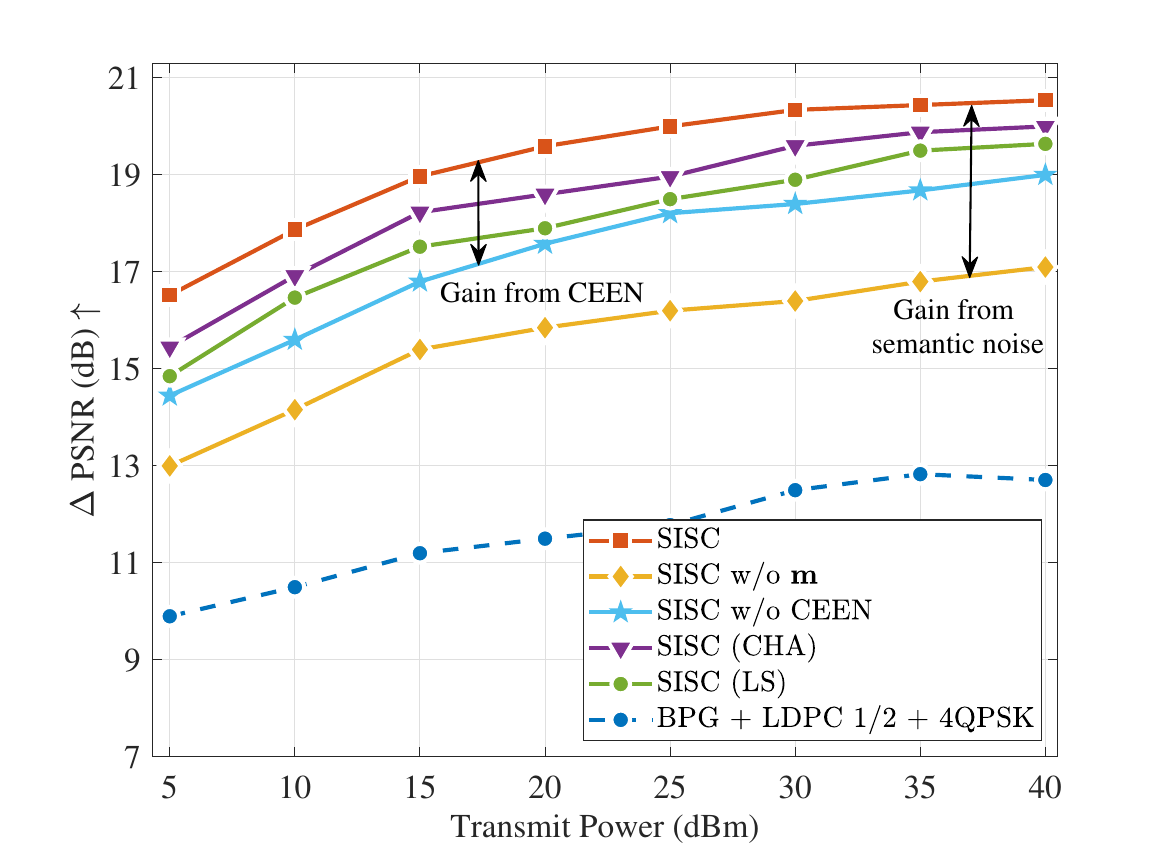}}
	\caption{Performance of ablation schemes to validate the effectiveness of the proposed semantic noise method and CEEN.}
	\label{Ablation_CE}
	\vspace{-0.25cm}
\end{figure}

	\subsubsection{Semantic Noise Analysis}
	As shown in Fig. \ref{Learnable_m}, we evaluate our learnable semantic noise generation mechanism under three fixed noise levels, and its PSNR gains are compared with those of the scheme without semantic noise. The annotations in the figure denote the semantic noise value changing with different powers. 
	The learnable mechanism adaptively decreases the semantic noise value $\mathbf{m}$ as the transmit power increases. 
	This result indicates that under high transmit power, the channel condition becomes more favorable for secure image transmission to the SU. In this case, only a slight semantic disturbance is required to maintain robustness while preserving semantic fidelity and stable PSNR performance across power levels.
	In contrast, under low transmit power conditions, the proportion of less important yet noise-sensitive elements increases significantly, thereby requiring a higher semantic noise level. Overall, the proposed SISC scheme effectively balances semantic robustness and fidelity across different channel conditions, yielding consistent PSNR gains over fixed semantic noise values, which further demonstrate the effectiveness of the proposed learnable semantic noise generation strategy.

\begin{figure*}[t]
	\centering
	\setlength{\belowcaptionskip}{-4.84em} 
	\subfloat{
		\includegraphics[width=0.965\linewidth]{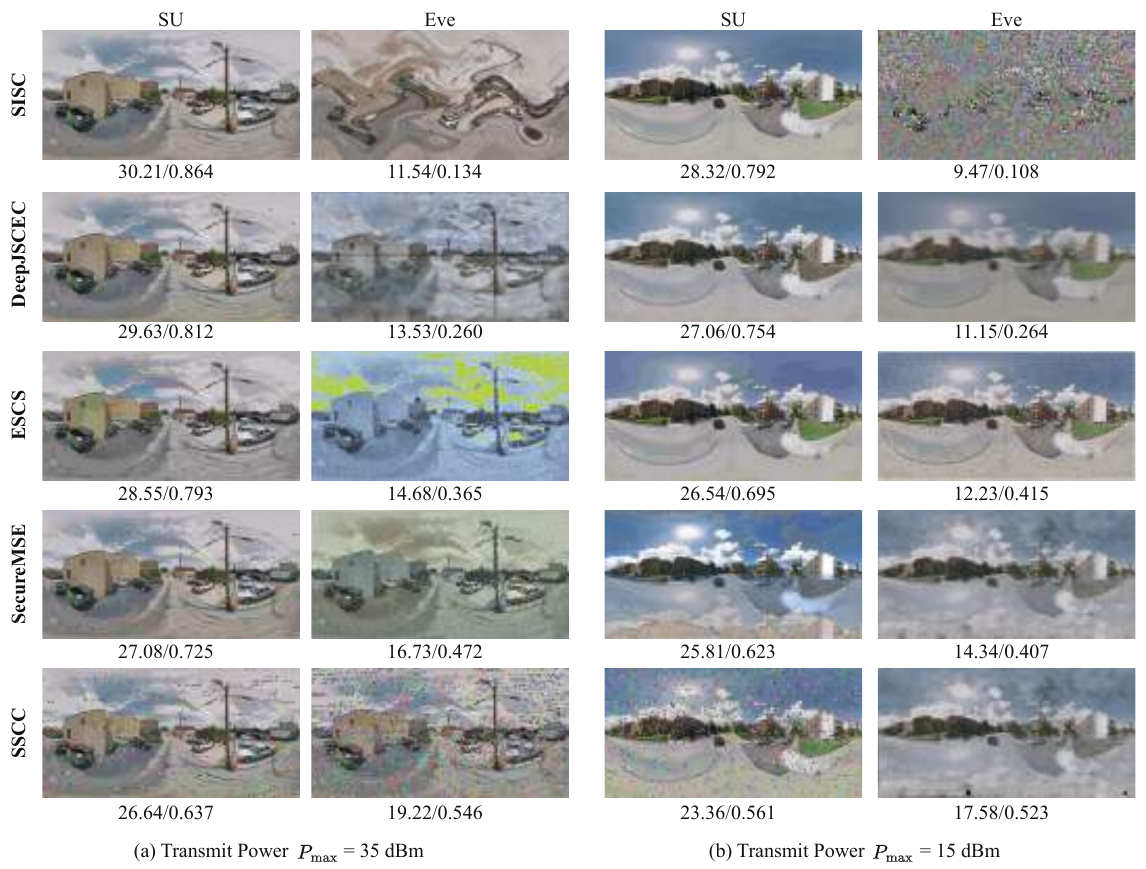}}
	\caption{Visualization of reconstructed images obtained by different schemes. The values below the images represent the PSNR(dB)/MS-SSIM results. The first row is the original image awhile the second to sixth columns show the reconstructions of the SISC, DeepJSCEC, ESCS, SecureMSE, BPG+LDPC3/4+4QPSK of SCSC, respectively. 
		(a) The first and second columns present the images recovered by the SU and Eve at $P_\mathrm{max}=35$ dBm, respectively. (b) The reconstruction results at $P_\mathrm{max}=15$ dBm.}
	\label{visual}
	\vspace{-0.35cm}
\end{figure*}

\subsubsection{Simulations for Different MIMO Modes}
We evaluate the proposed SISC scheme under various MIMO configurations and present simulation results with different antenna numbers. For a fair comparison, multiple sets of MIMO fading channel data are generated using identical parameters, with the exception of the transceiver antenna configurations.
The results in Table \ref{MIMO} show that increasing the number of antennas consistently improves SU reconstruction quality due to the enhanced spatial diversity and beamforming capability. More importantly, the performance gap between SU and Eve becomes even larger in more high-dimensional MIMO systems, indicating that the proposed security mechanism benefits from more spatial degrees of freedom. This is because the CSSCA-based optimization can exploit richer beamforming design space to better suppress Eve while maintaining SU performance.

	\subsection{Ablation Study}
	To further validate the effectiveness of our SISC network, the results of SISC without CEEN (``SISC w/o CEEN'') scheme and without SGN for generating semantic noise map (``SISC w/o $\mathbf m$'') scheme are shown in Fig. \ref{Ablation_CE} to demonstrate the benefits of the proposed CEEN and SGN modules, respectively.
	All of these schemes adopt the identical structure and configuration as SISC for the same network components.
	The performance metric here is defined as the PSNR difference between the SU and Eve, reflecting the security gain achieved by different schemes.
	The remaining components remain optimized without further fine-tuning.
	The performance of SISC w/o $\mathbf{m}$ degrades significantly compared with the other ablation schemes, indicating that the semantic noise map plays a crucial role in ensuring transmission security.
	We also conduct additional experiments on SISC employing different channel estimation strategies for comprehensive validation. Specifically, the schemes utilizing the conventional least squares (LS) estimator and the DL-based estimator in \cite{CHA} are denoted as ``SISC(LS)'' and ``SISC(CHA)'', respectively. 
	We can observe that SISC with the proposed CEEN outperforms SISC(CHA) and SISC(LS), which indicates the effectiveness of CEEN for accurate MIMO CSI acquisition.
	In this way, the proposed modules are verified to be effective during the wireless MIMO image secure transmission.

	\subsection{Visualization Results}
	To intuitively demonstrate the effectiveness of our proposed SISC, we further provide examples of transmitted and reconstructed images in Fig. \ref{visual}.
	The visualization results show that the SISC scheme outperforms other DL-based baseline methods in both image quality at the SU and resistance to eavesdropping. At the SU, reconstructed images at SISC preserve high-frequency object details and textures, with clearer edges and structures, due to the semantic-aware encoder-decoder and learnable semantic noise guided transmission. Higher transmit power further improves reconstruction quality, which validates the effectiveness of the transceiver beamformer optimization algorithm. When it comes to the traditional separated coding scheme, SISC obviously outperforms SCSC. In contrast, Eve’s reconstructions are blurred and structurally impaired. This ensures high-quality SU recovery while suppressing Eve.

	\subsection{Model Complexity Analysis}
	Table \ref{table} presents the number of model parameters, computational complexity, and throughput in terms of four frameworks.
	It can be observed that with similar parameters, SISC shows competitive performance compared to ECSC in terms of FLOPs and throughputs. 
	The DeepJSCEC framework exhibits lower computational cost and faster inference speed than SISC, owing to its compact auto-encoder structure.
	Even with semantic noise generation adopted, our SISC method does not incur large complexity but greatly protects the image secure transmission for SU.

	\begin{table}[t]
		\centering
		\caption{Complexity Comparison of Different Schemes.}
		\begin{tabular}{c|c|c|c}
			\hline
			Method & \makecell{Parameters(M)} & \makecell{FLOPs(G)} & \makecell{Throughput} \\
			\hline
			SISC & 59.47 & 14.92 & 157.2 \\
			DeepJSCEC & 43.61 & 12.45 & 266.3 \\
			ESCS & 60.08 & 16.38 & 141.7 \\
			SecureMSE & 36.16 & 8.26 & 122.5 \\
			\hline
		\end{tabular}
		\label{table}
		\vspace{-0.14cm}
	\end{table}

	\section{Conclusion}
	\label{sec:Conclusion}
	In this paper, we proposed a secure semantic image transmission framework named SISC for enhancing image reconstruction at the intended legitimate SU while impairing the recovery quality at the eavesdropper.
	Specifically, the beneficial semantic noise was designed by a semantic noise generation network, SNG, which mainly relies on the conditional variational autoencoder. 
	Furthermore, the CSSCA optimization method was employed to jointly optimize the transceiver beamformers, thereby concentrating signal energy toward the SU and suppressing information leakage to Eve.
	Experimental results demonstrated that the proposed SISC framework achieves superior performance in both image reconstruction quality and transmission security compared with existing benchmarks.
	Furthermore, SISC is a flexible framework capable of accommodating varying channels. In addition to robust image semantic transmission, the proposed framework exhibits strong scalability and can be extended to multi-modal wireless communication systems, such as text, audio, and video transmission.

\end{document}